\documentclass[fleqn,usenatbib]{mnras}
\usepackage[T1]{fontenc}
\usepackage{ae,aecompl}
\usepackage{graphicx}
\usepackage{epstopdf}
\usepackage{amsmath}
\usepackage{amssymb}
\usepackage{enumerate}
\usepackage{mathabx}
\usepackage{subfigure}
\usepackage{booktabs}
\usepackage{tablefootnote}
\usepackage{threeparttable}
\usepackage{color}

\pubyear{2020}
\begin{document}

\def\func#1{\mathop{\rm #1}\nolimits}
\def\unit#1{\mathord{\thinspace\rm #1}}

\title[The spin of MAXI J1820+070]{Physical origin of the nonphysical spin evolution of MAXI J1820+070}

\author[Guan et al.]{ J. Guan$^{1}$\thanks{%
E-mail: jguan@ihep.ac.cn}, L. Tao$^{1}$\thanks{%
E-mail: taolian@ihep.ac.cn}, J. L. Qu$^{1,2}$, S. N. Zhang$^{1,2}$, W. Zhang$^{1,2}$, S. Zhang$^{1}$, R. C. Ma$^{1,2}$, M. Y. Ge$^{1}$, \newauthor L. M. Song$^{1,2}$, F. J. Lu$^{1}$, T. P. Li$^{1,3}$, Y. P. Xu$^{1}$, Y. Chen$^{1}$, X. L. Cao$^{1}$, C. Z. Liu$^{1}$, L. Zhang$^{4}$, \newauthor Y. N. Wang$^{4}$, Y. P. Chen$^{1}$, Q. C. Bu$^{1}$, C. Cai$^{1,2}$, Z. Chang$^{1}$, L. Chen$^{5}$, T. X. Chen$^{1}$, Y. B. Chen$^{3}$, \newauthor W. W. Cui$^{1}$, Y. Y. Du$^{1}$, G. H. Gao$^{1,2}$, H. Gao$^{1,2}$, Y. D. Gu$^{1}$, C. C. Guo$^{1,2}$, D. W. Han$^{1}$, Y. Huang$^{1}$, \newauthor J. Huo$^{1}$, S. M. Jia$^{1}$, W. C. Jiang$^{1}$, J. Jin$^{1}$, L. D. Kong$^{1}$, B. Li$^{1}$, C. K. Li$^{1}$, G. Li$^{1}$, W. Li$^{1}$, X. Li$^{1}$, \newauthor X. B. Li$^{1}$, X. F. Li$^{1}$, Z. W. Li$^{1}$, X. H. Liang$^{1}$, J. Y. Liao$^{1}$, B. S. Liu$^{1}$, H. W. Liu$^{1}$, H. X. Liu$^{1}$, \newauthor X. J. Liu$^{1}$, X. F. Lu$^{1}$, Q. Luo$^{1,2}$, T. Luo$^{1}$, X. Ma$^{1}$, B. Meng$^{1}$, Y. Nang$^{1,2}$, J. Y. Nie$^{1}$, G. Ou$^{1}$, \newauthor X. Q. Ren$^{1,2}$, N. Sai$^{1,2}$, X. Y. Song$^{1}$, L. Sun$^{1}$, Y. Tan$^{1}$, C. Wang$^{2,6}$, L. J. Wang$^{1}$, P. J. Wang$^{1,2}$, \newauthor W. S. Wang$^{1}$, Y. S. Wang$^{1}$, X. Y. Wen$^{1}$, B. B. Wu$^{1}$, B. Y. Wu$^{1,2}$, M. Wu$^{1}$, G. C. Xiao$^{1}$, S. Xiao$^{1,2}$, \newauthor S. L. Xiong$^{1}$, R. J. Yang$^{7}$, S. Yang$^{1}$, Y. J. Yang$^{1}$, Y. J. Yang$^{1}$, Q. B. Yi$^{1,2}$, Q. Q. Yin$^{1}$, Y. You$^{1}$, \newauthor F. Zhang$^{1}$, H. M. Zhang$^{1}$, J. Zhang$^{1}$, P. Zhang$^{1}$, W. C. Zhang$^{1}$, Y. F. Zhang$^{1}$, Y. H. Zhang$^{1,2}$, \newauthor H. S. Zhao$^{1}$, X. F. Zhao$^{1,2}$, S. J. Zheng$^{1}$, Y. G. Zheng$^{1,7}$, D. K. Zhou$^{1,2}$
\\
$^{1}$ Key Laboratory of Particle Astrophysics, Institute of High Energy
Physics, Chinese Academy of Sciences, Beijing 100049, China\\
$^{2}$ University of Chinese Academy of Sciences, Chinese Academy of Sciences, Beijing 100049, China\\
$^{3}$ Department of Astronomy, Tsinghua University, Beijing 100084, China\\
$^{4}$ Physics and Astronomy, University of Southampton, Southampton, Hampshire SO17 1BJ, UK\\
$^{5}$ Department of Astronomy, Beijing Normal University, Beijing 100088, China\\
$^{6}$ Key Laboratory of Space Astronomy and Technology, National Astronomical Observatories, Chinese Academy of Sciences,\\ 
Beijing 100012, China\\
$^{7}$ College of physics Sciences \& Technology, Hebei University, No. 180 Wusi Dong Road, Lian Chi District, Baoding City,\\ 
Hebei Province 071002, China}
\date{Accepted XXX. Received YYY; in original form ZZZ}
\maketitle

\begin{abstract}
    We report on the \emph{Insight}-HXMT observations of the new black hole X-ray binary MAXI J1820+070 during its 2018 outburst. Detailed spectral analysis via the continuum fitting method shows an evolution of the inferred spin during its high soft sate. Moreover, the hardness ratio, the non-thermal luminosity and the reflection fraction also undergo an evolution, exactly coincident to the period when the inferred spin transition takes place. The unphysical evolution of the spin is attributed to the evolution of the inner disc, which is caused by the collapse of a hot corona due to condensation mechanism or may be related to the deceleration of a jet-like corona. The studies of the inner disc radius and the relation between the disc luminosity and the inner disc radius suggest that, only at a particular epoch, did the inner edge of the disc reach the innermost stable circular orbit and the spin measurement is reliable. We then constrain the spin of MAXI J1820+070 to be $a_*=0.2^{+0.2}_{-0.3}$. Such a slowly spinning black hole possessing a strong jet suggests that its jet activity is driven mainly by the accretion disc rather than by the black hole spin.

\end{abstract}

\label{firstpage} \pagerange{\pageref{firstpage}--\pageref{lastpage}}


\section{Introduction}
\label{sec:intro}

An astrophysical black hole (BH) is completely described by two parameters: its mass and spin \citep{Shafee2006}. While the mass supplies a scale, the spin changes the geometry of the space-time and the ways a BH interacting with its surrounding environment, and thus is of great importance to understanding the BH physics \citep{McClintock2007}. For instance, sufficient data on BH spins is needed to better understand whether the relativistic jets are driven by Blandford-Znajek (BZ) mechanism \citep{BZ1977} or Blandford-Payne (BP) mechanism \citep{BP1982}. Understanding BH formation and BH X-ray binary evolution also require knowledge of BH spin. Evidence for natal extreme spins provides strong support for the collapsar models of long gamma-ray bursts. Spin is also crucial to models of the characteristic low-frequency quasi-periodic oscillations (LFQPOs) of accreting BHs and the gravitational wave astronomy in improving the waveform of two inspiralling BHs \citep[][and references therein]{McClintock2006, Steiner2011}.

However, unlike the mass that can be relatively straightforward obtained by dynamical studies, the spin is much harder to be measured. Only at the end of last century, great breakthroughs have been made in deriving credible measurement of BH spin via two independent methods: (1) the Fe K$\alpha$ method that models the profile of the reflection-fluorescent features in the disc, especially the relativistically-broadened and asymmetric iron line \citep{Fabian1989, Reynolds2003}; (2) the continuum-fitting (CF) method that fits the thermal X-ray continuum spectrum from the disc to the Novikov-Thorne \citep[NT,][]{Novikov1973} thin disk model \citep{Zhang1997}. The spin measurements derived from the two methods 
have been compared and discussed in detail \citep{Reynolds2020,Salvesen2021}.

For both methods, a fundamental assumption is that the accretion disc is extended to the innermost stable circular orbit (ISCO). This link between the inner disk radius ($R_{\rm in}$) and that of the ISCO ($R_{\rm ISCO}$) is strongly supported by theoretical simulations \citep{Shafee2008, Penna2010} and empirical evidence that $R_{\rm in}$ does not change during the high soft (HS) state \citep{Narayan2008, Steiner2010, Kulkarni2011}. An interpretation of a thermal disc with a constant $R_{\rm in}$ is provided by the relation between the intrinsic disc luminosity and the temperature, which follows the expected $L_{\rm disk} \propto T_{\rm in}^4$ \citep{McClintock2006}. Since $R_{\rm ISCO}/(GM/c^2)$ is a monotonic function of the BH dimensionless spin parameter $a_{*}$ ($a_{*}=cJ/GM^{2}$, where $M$ and $J$ are the BH mass and angular momentum, respectively); knowing $R_{\rm in}$, and thereby $R_{\rm ISCO}$, allows one to directly derive the BH spin $a_{*}$. In the CF method, one determines $R_{\rm in}$ and then $a_{*}$ via measurements of the X-ray spectral shape and luminosity of the disc emission, since the spin will influence the gravitational well of the BH, leading to the changes of the hardness and efficiency for converting the accreted rest mass into radiated energy \citep{McClintock2006}.

It is worth to note that the boundary between the intermediate state and the HS state is difficult to distinguish and whether the disc is truncated in the intermediate state is still debated; it is thus challenging to determine when the disc indeed extends to the ISCO in the intermediate state. Detailed studies of the spectral evolution may provide the most feasible way. No matter whether the state transition from the intermediate state to  the HS state is driven by the shrinking of the inner accretion disc \citep{Plant2014, Ingram2011} or by the collapse of the corona \citep{Garcia2015, Kara2019}, we may expect a spectral softening and change of the reflection fraction. Moreover, for the former case, the inner disc radius should decrease and the disk luminosity would deviate from the relation of $L_{\rm disk} \propto T_{\rm in}^4$, which means that the disc is still truncated and has not reached the BH's ISCO. While for the latter case, the Compton component would weaken, but the inner disc has already reached the ISCO. In other words, only if the inner disc radius is stable and the disc luminosity follows $L_{\rm disk} \propto T_{\rm in}^4$, the disc has extended to the ISCO and thus a reliable measurement of the spin can be obtained. 

Besides selecting spectra when $R_{\rm in}$ has reached $R_{\rm ISCO}$, in order to derive credible $a_{*}$ via the CF method, it is also essential to (1) restrict to luminosity below $30\%$ of the Eddington limit in order to make sure the application of the thin disk approximation; (2) have accurate measurements of the BH mass $M$, distance $D$, and inclination of the accretion disk \citep{McClintock2007}. The robustness of the CF method has been confirmed in many stellar-mass BHs, giving spin values ranging from small \citep[$a_{*}\approx0.1$,][]{Gou2010}, moderate \citep[$a_{*}\approx0.49-0.85$,][]{Shafee2006, Steiner2011} to extreme \citep[$a_{*}>0.98$,][]{McClintock2006, Gou2011, Gou2014, Zhao2020a}. 

MAXI J1820+070 (hereafter ‘MAXI J1820’) is a new Galactic black hole candidate discovered by MAXI on 2018 March 11 \citep{Kawamuro2018}. Soon it was identified with ASASSN-18ey \citep{Tucker2018} detected five days earlier by the All-sky Automated survey for supernovae \citep{Denisenko2018}. This luminous X-ray source, brighter than 4\,Crab, making itself one of the brightest X-ray transients \citep{Fabian2020}. It has triggered vast multi-wavelength studies, revealing plentiful phenomena. Fast variability, QPOs, powerful flares, radio jets as well as low linear polarisation have been found from the source \citep[][and references therein]{Veledina2019,Wang2020}. For instance, strong observational evidence has been presented for linking the appearance of type-B QPOs and the launch of discrete jet ejections \citep{Homan2020}. The discovery of LFQPOs above 200\,keV in this source makes it unique, since it is the highest energy LFQPO detected in any BH binary known so far \citep{Ma2020}. Its hard-soft-hard behaviour and X-ray reverberation lags are consistent with an accreting black hole \citep{Fabian2020}. The optical counterpart of MAXI J1820 is comprehensively studied by \cite{Torres2019,Torres2020}. The measurement of a mass function $f(M)=5.18\pm0.15\,M_{\odot}$ immediately established MAXI J1820 as a dynamically confirmed BH binary hosting a BH of $8.48^{+0.79}_{-0.72}\,M_{\odot}$. A precise measurement of the radio parallax of MAXI J1820 using VLBA and VLBI has provided a model-independent distance of $2.96\pm0.33$\,kpc and a jet inclination angle of $63\pm3^{\circ}$ \citep{Atri2020}. Thus, the three key quantities that are essential for determining the spin of the BH via the CF method are available. However, few specific measurements for MAXI J1820’s spin have been performed, except that \citet{Atri2020} has suggested the BH in MAXI J1820 is likely slowly spinning, and \citet{Buisson2019} favors a low to moderate spin BH, and \citet{Fabian2020} gives a relation curve between the spin and the inclination.

The Hard X-ray Modulation Telescope, dubbed as \emph{Insight}-HXMT \citep{Zhang2020}, also carried out Target of Opportunity (ToO) observations on this source, which have covered the whole HS state with an effective exposure of $\sim500$ ks. Benefited from the large effective area, the broad energy coverage and being free from pipe-up effect, \emph{Insight}-HXMT could simultaneously constrain the soft thermal component, the Compton power-law and the reflection components. Thus it is quite ideal to study the spin of MAXI J1820 with \emph{Insight}-HXMT via the CF method. 

The paper is organized as follows. In Section \ref{sec:data}, we describe the observations and data reduction, and in Section \ref{sec:results} the applied spectral models and results, as well as an exploration of the systematic uncertainties inherent to the CF method. Presented in Section \ref{sec:dis} is the discussion. We offer our conclusions in the final section.

\section{Observations and data reduction}

\label{sec:data}

\emph{Insight}-HXMT is the first Chinese X-ray astronomy satellite, which was successfully launched on 2017 June 15. It carries three slat-collimated instruments on board: the low energy X-ray telescope \citep[LE, $1-15$\,keV, 384\,$\rm{cm}^{2}$,][]{Chen2020}, the medium energy X-ray telescope \citep[ME, $5-35$\,keV, 952\,$\rm{cm}^{2}$,][]{Cao2020}, and the high energy X-ray telescope \citep[HE, $20-250$\,keV, 5100\,$\rm{cm}^{2}$,][]{Liu2020}. The entire outburst of MAXI J1820 has been observed with \emph{Insight}-HXMT between 2018 March and October. In this paper, we perform the spectral analysis of MAXI J1820 during its HS state (from MJD 58310 to MJD 58380) with 49 \emph{Insight}-HXMT observations. The detailed information for these observations are listed in Table \ref{tbl:Observations}.

We extract the data using the \emph{Insight}-HXMT Data Analysis software (HXMTDAS) v2.01. The data are filtered with the following criteria recommended by the \emph{Insight}-HXMT team: (1) the offset for the pointing position is $\le0.05^{\circ}$; (2) the elevation angle is $>6^{\circ}$; (3) the geomagnetic cutoff rigidity is $>8$; (4) the extraction time is at least $300$\,s before or after the South Atlantic Anomaly passage. To avoid possible contamination from nearby sources and the bright Earth, only detectors with small field of view (FOV) are applied. We focus on data obtained by LE and ME, given that the net photons obtained by HE have low statistics during the HS state and that the other two instruments have already provided adequate energy coverage. The backgrounds for LE and ME are estimated by the aid of the blind detectors, given that the spectral shapes of the particle backgrounds are the same for both the blind and small FOV detectors and that the correction factor of their intensities can be calibrated using blank sky observations \citep{liao2020, guo2020}. The energy bands adopted for spectral analysis are listed in Table~\ref{tbl:models}. For LE, the spectra are rebinned to have at least 100 counts per bin, while for the ME spectra before and after $21$\,keV, every 2 and 5 channels are rebinned into one bin, respectively. 
 
\section{Results}

\label{sec:results}

\subsection{Light curve and hardness ratio}
\begin{figure}
    \centering\includegraphics[width=0.48\textwidth]{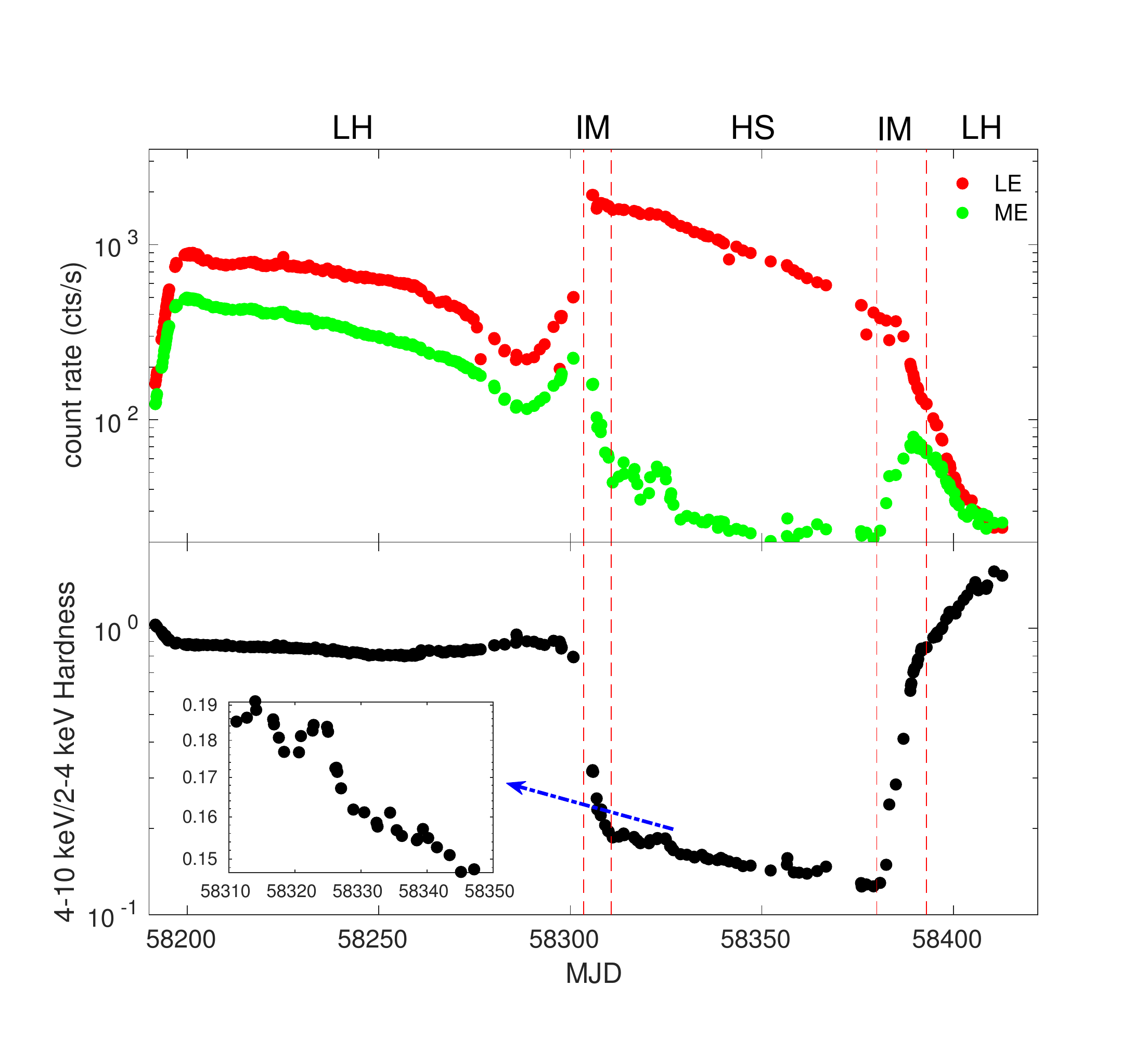}
    \caption{\textsl{Top}: Evolution of the X-ray count rate of LE (red dots) and ME (green dots). \textsl{Bottom}: the hardness ratio of MAXI J1820 defined as the ratio of the count rates of LE in the $4-10$\,keV to $2-4$\,keV bands. The red dashed lines mark the time of two intermediate states. The sudden drop of the hardness ratio at MJD$\sim$58330 is highlighted in the inset.}
    \label{fig:hardness}
\end{figure}

In Figure \ref{fig:hardness}  we show the light curves of the 2018 outburst of MAXI J1820 in the $2-35$\,keV band and the hardness ratio (defined as the ratio of the count rates in the $4-10$\,keV to $2-4$\,keV bands). The source goes through a clear evolution and is extremely variable during the whole outburst. Adopting the states defined in \citet{Shidatsu2019}, we focus on the spectra dominated by the thermal emission, i.e. the HS state, in which the hardness ratio satisfies the empirical selection criterion \citep[HR$<0.3$,][]{McClintock2006}. 

During the HS state, both the light curves and the hardness ratio decrease with time and an interesting jump (drop) at MJD $\sim58330$ is distinguishable in the hardness ratio and ME light curve, which might indicate the decrease of the hard component. The same situation has also been found in the NICER data, which showed a steeper drop in its hardness ratio \citep{Homan2020}.

\subsection{Spectral properties}
\label{sec:spectral properties}

The spectral fitting is carried out using the software package XSPEC V12.10.1. We adopt several phenomenological and physical models (M1-M4 in Table \ref{tbl:models}) to characterize the broad band spectra. All models include the interstellar absorption effect by implementing the TBABS model component with \citet{Wilms2000} abundances and \citet{Verner1996} cross sections. Because we ignore photons below 2\,keV due to the limitation of the calibration, which makes us unable to define the column density $N_{\rm H}$ well, we fix it to a well-study value of $0.16 \times 10^{22}\,\rm{cm}^{-2}$ \citep{Uttley2018,Bharali2019}. A CONSTANT component is included to reconcile the calibration discrepancies between LE and ME (throughout the paper, we fix the multiplicative factor of LE to 1). Moreover, the BH mass of $M=8.48^{+0.79}_{-0.72} \,M_{\odot}$ \citep{Torres2020}, the inclination of $i=63\pm3^{\circ}$ and the distance of $D=2.96\pm0.33$\,kpc \citep{Atri2020} are adopted. The fitting procedure minimizes the $\chi^2$ goodness-of-fit statistic. The uncertainty estimated for each spectral parameter is at $90\%$ confidence level, unless otherwise stated, and a systematic error of $1.5\%$ is added in the fitting. 

We start our analysis by modelling the $2-5$\,keV (LE) and $10-20$\,keV (ME) spectra with non-reflection models, i.e. models M1-M3. The $5-10$\,keV of LE has been ignored because a weak extra structure lying there is still debated \citep{Fabian2020}. The $20-35$\,keV band of ME is discarded as it is the main contributing energy band of the Compton hump. Then we analyze the full band ($2-35$\,keV) spectra with a more sophisticated model (model M4) that attributes the extra structure to reflection features in order to test the impact on the spin measurement. We note that the origin of the structure may require in-depth investigations, but is beyond the scope of this work.

First, we fit the spectrum with an absorbed multicolour disc blackbody \citep[DISKBB,][]{Mitsuda1984, Makishima1986} plus a power-law component (M1). The total set-up of model M1 (Table~\ref{tbl:models}) is: CONSTANT*TBABS*(DISKBB+POWERLAW). As we only use part of the energy band, we constrain the CONSTANT of ME to vary from $0.9$ to $1.1$. We also have tried to fix it at 1 and find that the fitting results are barely affected. Model M1 provides a good description of the spectra with the reduced $\chi^2$ ranging from $0.62$ to $0.99$. 

\begin{figure}
    \centering\includegraphics[width=0.48\textwidth]{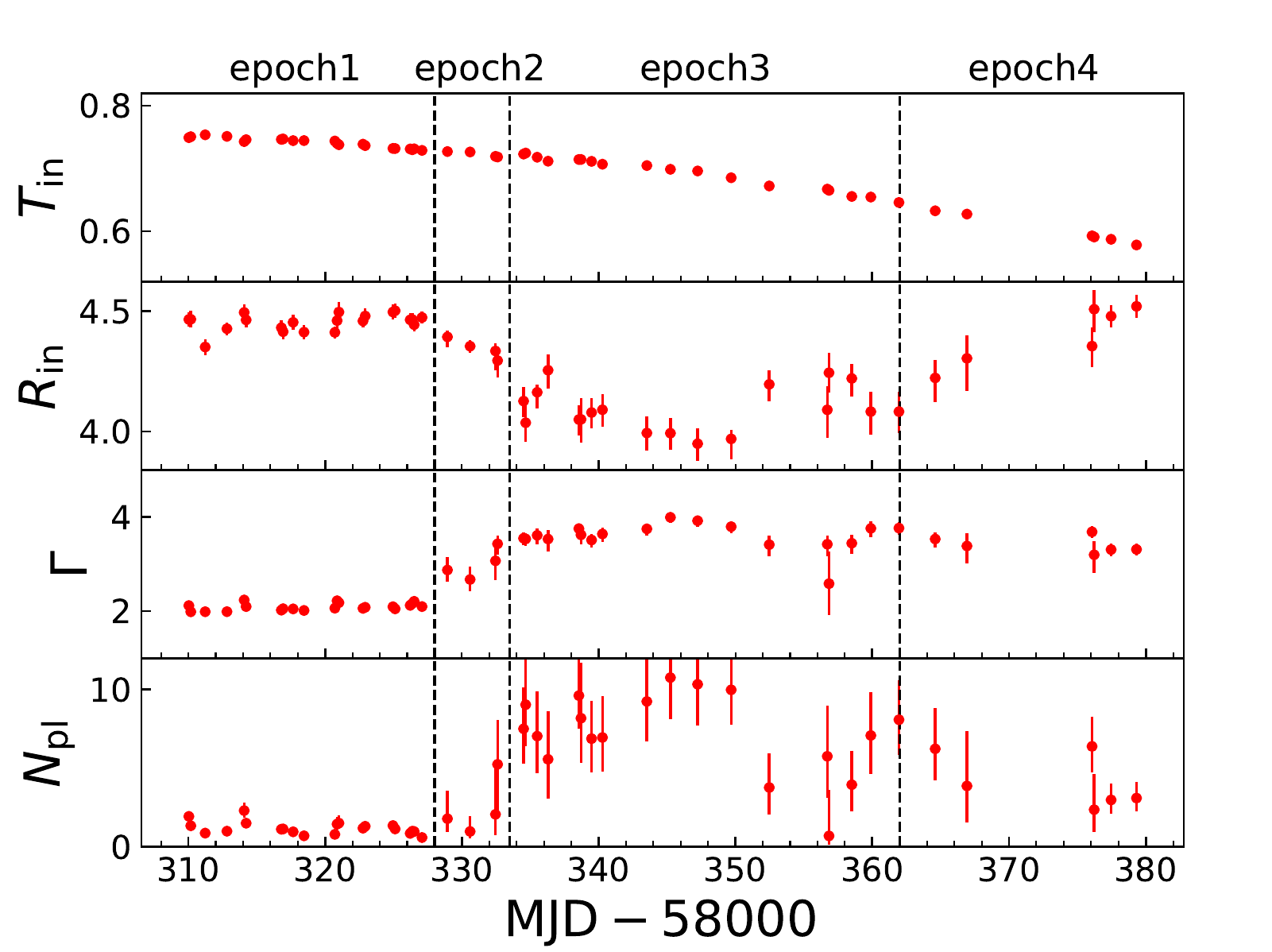}
    \caption{Evolultion of the spectral parameters of M1. $T_{\rm in}$ is the inner disc temperature; $R_{\rm in}$ is the apparent inner disc radius in units of $R_{\rm g}$, calculated from the normalization of DISKBB, $M$, $i$ and $D$; $\Gamma$ is the photon index; $N_{\rm pl}$ is the normalization of POWERLAW.}
    \label{fig:diskbb}
\end{figure}

\begin{figure}
    \centering\includegraphics[width=0.48\textwidth]{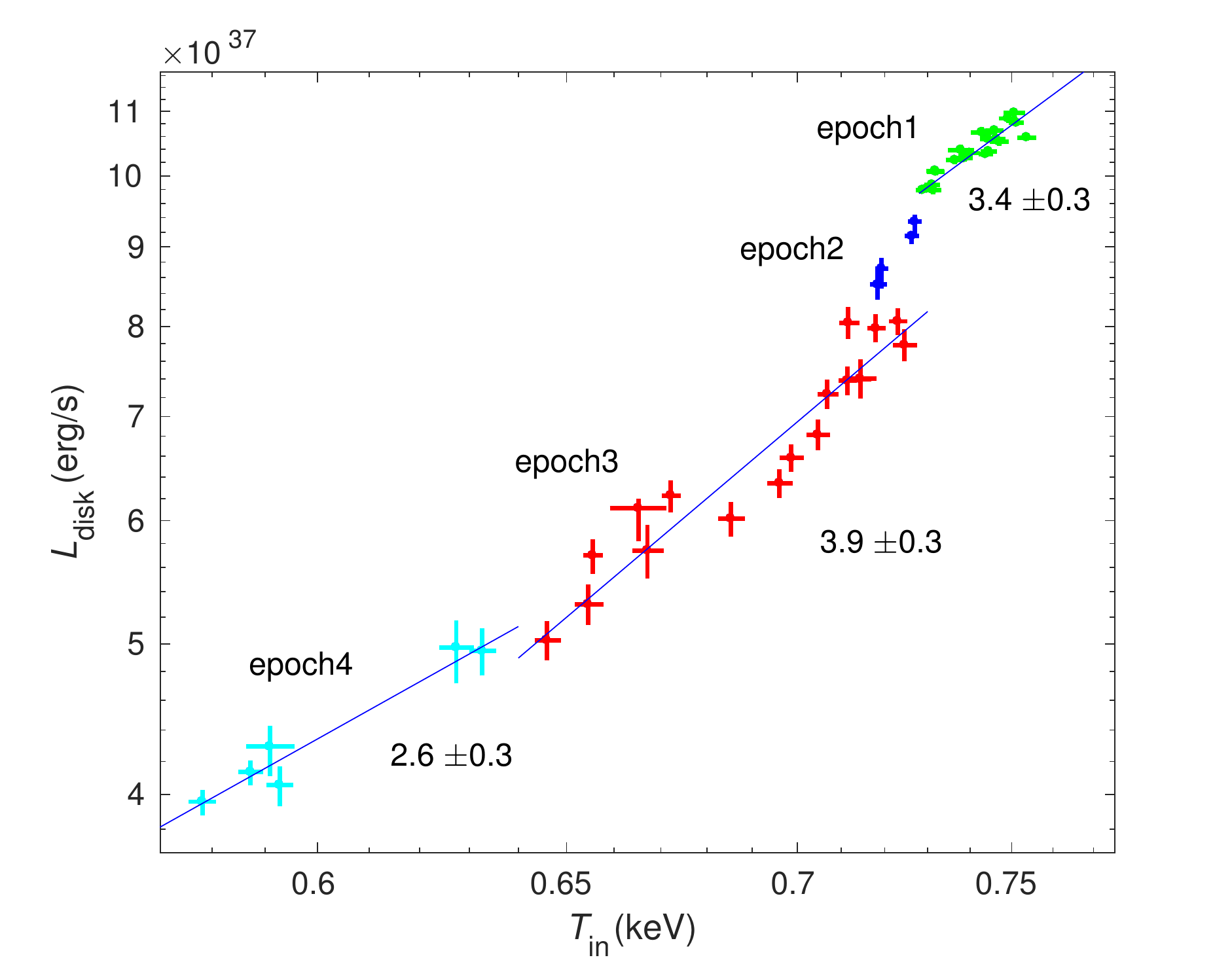}
    \caption{The disc luminosity ($L_{\rm disk}$) versus the disc inner temperature ($T_{\rm in}$). Different epochs are marked with different colors. The blue lines represent the best-fitting relations.}
    \label{fig:LT4}
\end{figure}

The evolution of the spectral parameters is shown in Figure~\ref{fig:diskbb} and Table~\ref{tbl:M1}. The inner disc temperature $T_{\rm in}$ shows a small decline, with the central values ranging from 0.753 to 0.578  keV. Using $M=8.48^{+0.79}_{-0.72} \,M_{\odot}$, $i=63\pm3^{\circ}$ and $D=2.96\pm0.33$\,kpc, the normalization of DISKBB is converted to the apparent disc radius $R_{\rm in}$ in units of $R_{\rm g}$, which shows a dramatic evolution with time. It remains stable at $\sim4.5$\,$R_{\rm g}$ till MJD $\sim58330$ then drops to a new stable level of $\sim4$\,$R_{\rm g}$ till MJD $\sim58360$ and finally increases back to $\sim4.5$\,$R_{\rm g}$. The photon index and the normalization of POWERlAW shows an opposite evolution trend with that of $R_{\rm in}$. Motivated by the evolution of the spectral parameters, we plot the relation of the disc luminosity $L_{\rm disk}$ and the inner disc temperature $T_{\rm in}$. In principle, it is only when the inner disc reach the ISCO ($L_{\rm disk}$ and $T_{\rm in}$ follows $L_{\rm disk}\propto T_{\rm in}^4$) that the spin measurement via CF method is valid. Thus we need to test whether and when the inner disc reaches the ISCO. Figure~\ref{fig:LT4} plots the relation between $L_{\rm disk}$ and $T_{\rm in}$, which could roughly be divided into four groups according to the switch time of the evolution trend of $R_{\rm in}$. The green points in Figure \ref{fig:LT4} belong to the first epoch (from MJD $58310$ to  MJD $\sim58330$) and the blue ones correspond to the jumping time (from MJD $\sim58330$ to  MJD $\sim58333$), while the red and cyan  points denote the third (from MJD $\sim58333$ to  MJD $\sim58360$) and fourth (from MJD $\sim58360$ to  MJD $58380$) epoch, respectively. Setting aside the data corresponding to the jumping time (blue points in  Figure \ref{fig:LT4} ), we fit $L_{\rm disk}$ versus $T_{\rm in}$ in other three groups to a power-law function ($L_{\rm disk}\propto T_{\rm in}^\alpha$) separately and find that $\alpha$ is $3.4\pm0.3$, $3.9\pm0.3$, and $2.6\pm0.3$, respectively. Only the middle group satisfies the expected $L_{\rm disk}\propto T_{\rm in}^4$, indicating that only at this epoch did the inner disc reach the ISCO. It coincides with the evolution trend of $R_{\rm in}$, which reach the minimum at this epoch. 

As the next step, in order to derive the spin, we replace DISKBB with KERRBB2 \citep{McClintock2006} and replace POWERLAW with the empirical comptonization model SIMPL \citep{Steiner2009}. The total set-up of model M2 (Table \ref{tbl:models}) is: CONSTANT*TBABS*(SIMPL*KERRBB2). The hybrid code KERRBB2 is a modified version of KERRBB \citep{Li2005}. It inherits the special features of KERRBB, and incorporates the effects of spectral hardening via a pair of look-up tables for the spectral hardening factor $f$ corresponding to two values of the viscosity parameters: $\alpha=0.01, 0.1$ \citep{McClintock2006, Gou2011}. Throughout this work, we adopt $\alpha=0.1$ \citep{Steiner2011}. The look-up table is calculated by the aid of BHSPEC \citep{Davis2005}. Thus, the model KERRBB2 has just two fitting parameters: the spin $a_{*}$ and the mass accretion rate $\dot{M}$. We turn on the effects of the returning radiation and limb darkening, fix the torque at the inner boundary of the disk to zero and set the normalization to unity. Since SIMPL redistributes input photons to energies where the response matrices of \emph{Insight}-HXMT might be limited, we extend the sampled energies to $0.01-500$\,keV in XSPEC to adequately cover the relevant range \citep{Steiner2009}. 

\begin{figure}
    \centering\includegraphics[width=0.48\textwidth]{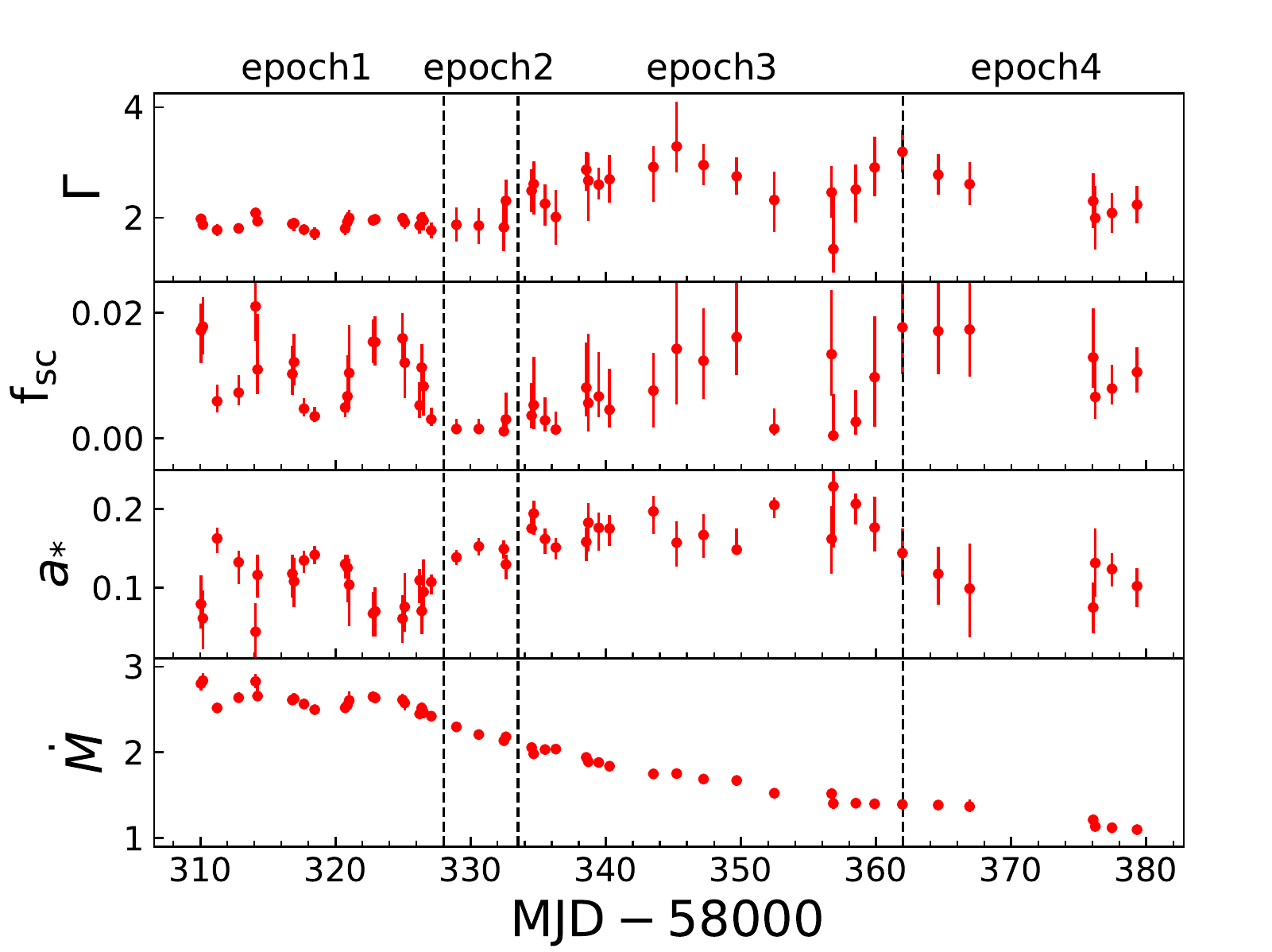}
    \caption{Evolution of the spectral parameters of M2. $\Gamma$ is the photon index; $f_{\rm sc}$ is the scattering fraction; $a_*$ is the spin; $\dot{M}$ is the mass accretion rate in units of $10^{18}$ \,g/s.}
    \label{fig:krrbbSimpl}
\end{figure}

\begin{figure}
    \centering\includegraphics[width=0.48\textwidth]{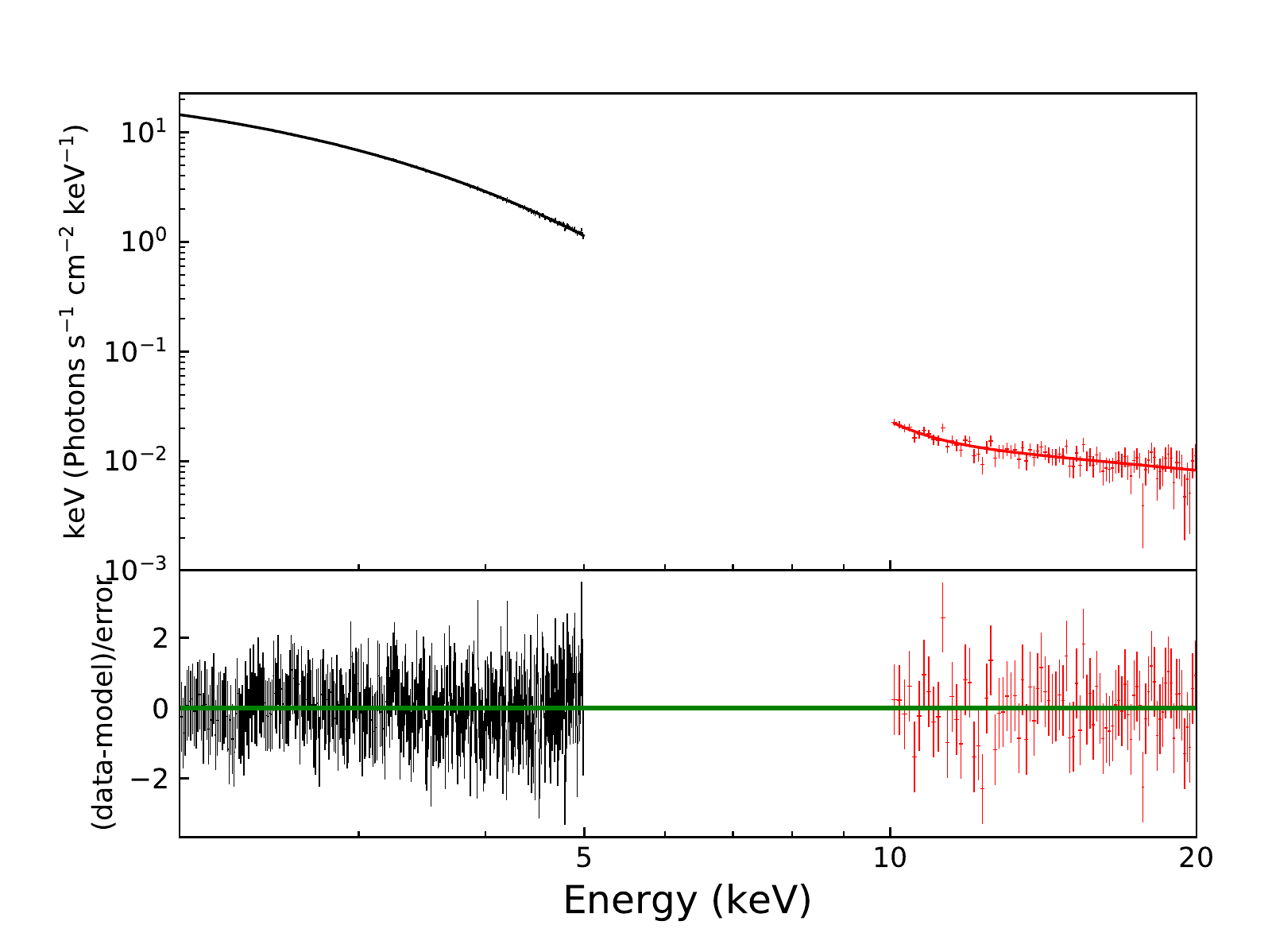}
    \caption{A representative spectrum (ObsID P011466110001) of MAXI J1820 fitted with M2.}
    \label{fig:krrbbSimplSpectrum}
\end{figure}

\begin{figure}
    \centering\includegraphics[width=0.48\textwidth]{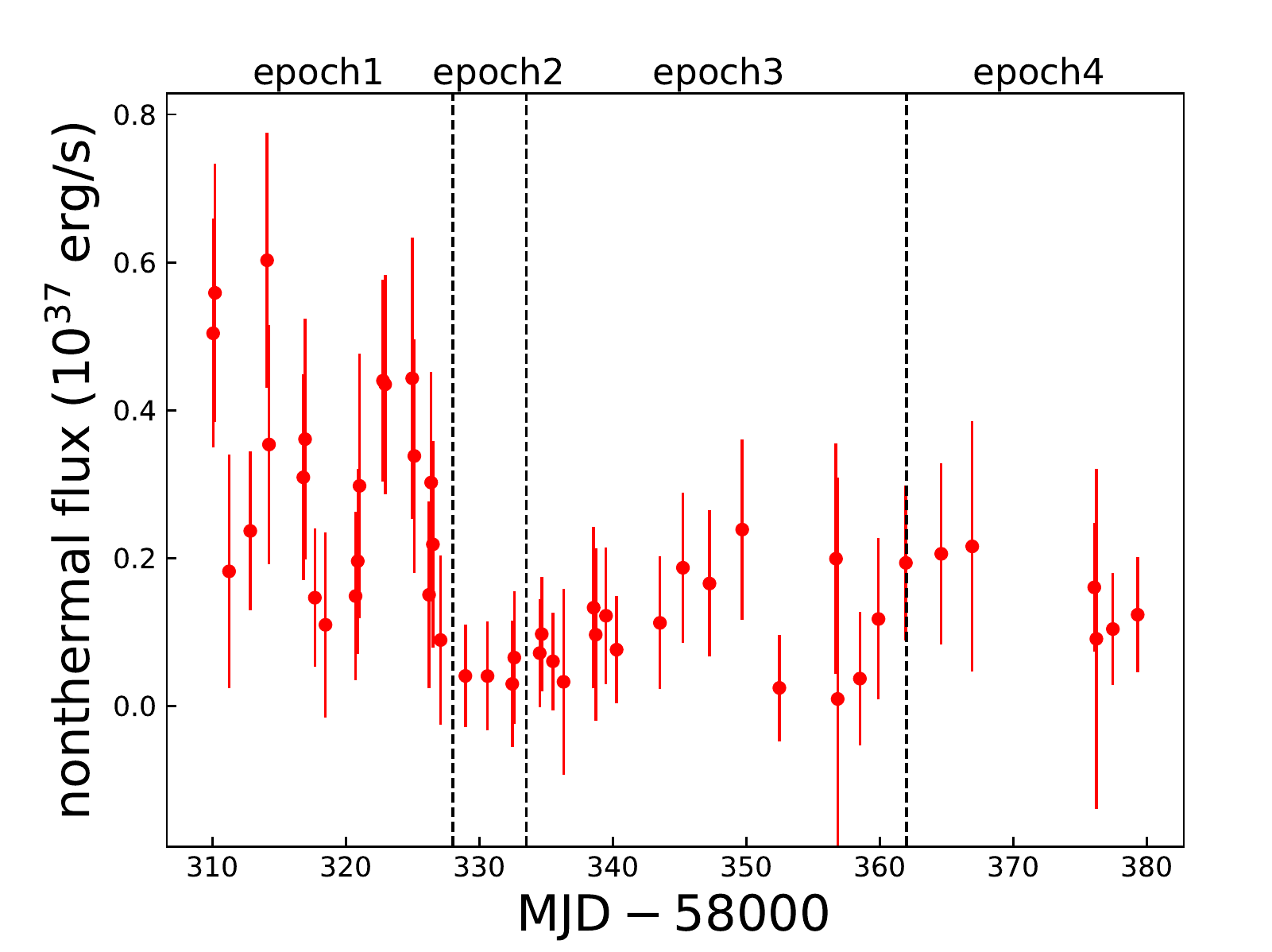}
    \caption{Evolution of the non-thermal luminosity.}
    \label{fig:nonthermal}
\end{figure}

The fitting results are shown in Figure~\ref{fig:krrbbSimpl} and Table~\ref{tbl:M2}. A representative plot of the fitting spectrum is given in Figure~\ref{fig:krrbbSimplSpectrum}. We obtain an opposite evolution trend of the spin compared with that of $R_{\rm in}$. The measured spin stays at $\sim0.1$ at the first epoch then increases to $\sim0.18$ at the third epoch then decreases. The $\dot{M}$ decreases slowly before MJD $\sim58330$ then follows by a rapid decline. The disk luminosity calculated with $L=\eta\dot{M}c^{2}$  also decreases, with an Eddington-scaled luminosity $l=L/L_{\rm Edd}$ ranging from 0.137 to 0.054 (Table \ref{tbl:M2}), satisfying the assumption of a thin disc. The evolution of the photon index in SIMPL is similar to that of POWERLAW in M1. Due to the statistic fluctuation, the evolution trend of $f_{\rm sc}$ is not so clear, however a drop at MJD $\sim58330$ is obvious. All $f_{\rm sc}$ are comparable with 0.025 (a typical value for the HS state), satisfying the selection criteria $f_{\rm sc}<25\%$ \citep{Steiner2011}. As $f_{\rm sc}$ is defined as the fraction of the seed photons supplied by KERRBB2 being scattered into the power-law tail, it represents the strength of the non-thermal component. To confirm this idea, the evolution of the non-thermal luminosity is plotted, which does drop at MJD$\sim$58330 (Figure~\ref{fig:nonthermal}), indicating an decrease of the non-thermal component. We have also tried to fix the CONSTANT of ME to 1 for all spectral fittings and find that the evolution trends of all parameters are seldom affected (indeed the errors of the parameters become smaller and the up-down-up evolution trend of $f_{\rm sc}$ is clearer). The spectral hardening factor $f$ is provided in Table~\ref{tbl:M2}, which decreases from $1.65$ to $1.57$. 

\begin{figure}
    \centering\includegraphics[width=0.5\textwidth]{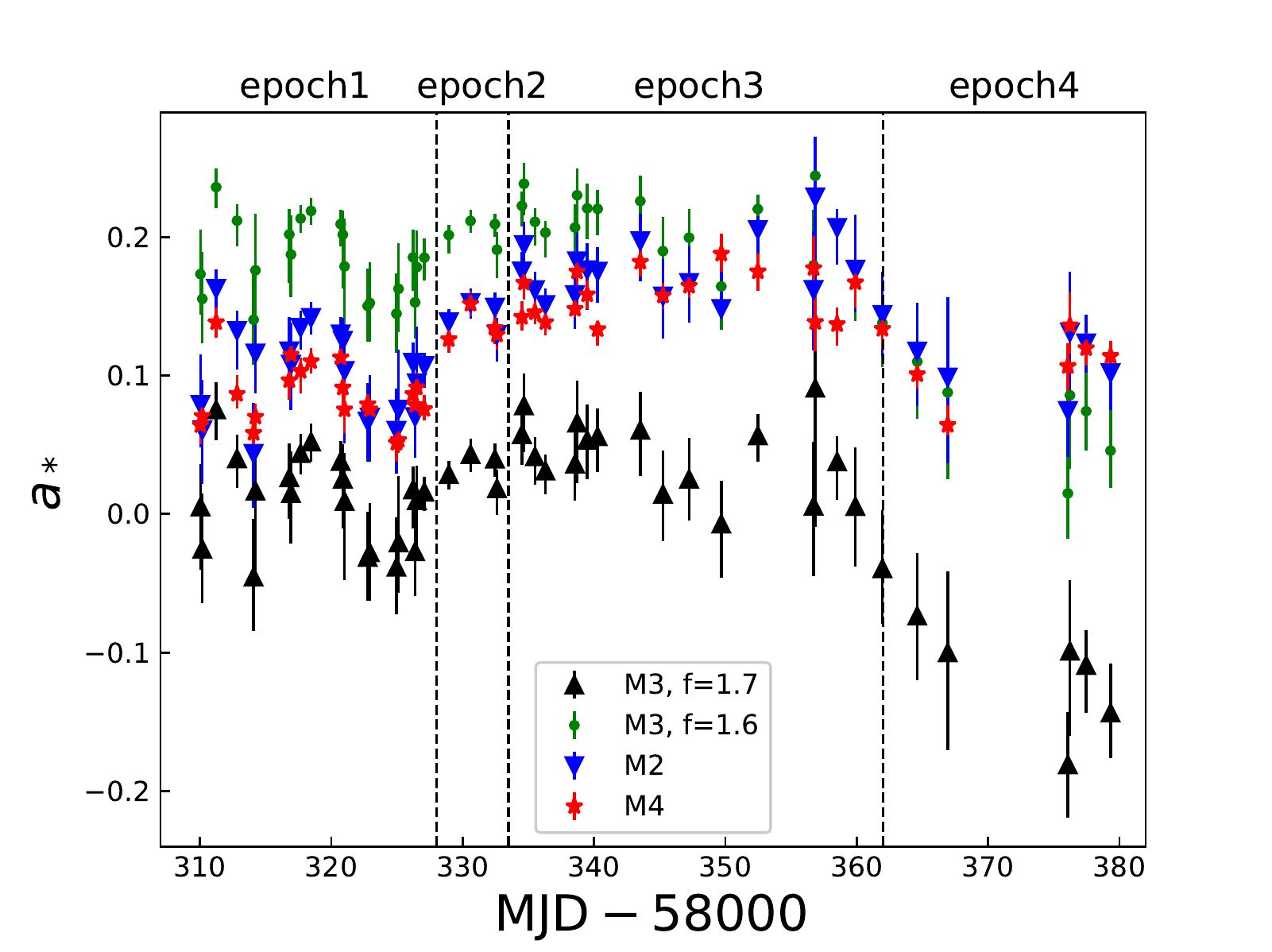}
    \caption{Evolultion of the spin corresponding to different models.}
    \label{fig:spin}
\end{figure}

For comparison, we derive the spin from a regressive disk model compared to KERRBB2, i.e. Model M3: CONSTANT*TBABS*(SIMPL*KERRBB). The spectral hardening factor $f$ of KERRBB is respectively fixed at $1.6$ and $1.7$ for all observations. The model yields a similar spin evolution trend as that of M2, though the jump at MJD $\sim58330$ is not so obvious (Figure \ref{fig:spin}).

Finally, we attempt to characterize the reflection component using the full 2--35\,keV spectra with a sophisticated model (M4: CONSTANT*TBABS*(SIMPLR*KERRBB2+KERRCONV
*(IREFLECT*SIMPLC))), to evaluate the impact on the spin measurement and understand the changes of the accretion flow and the interaction between the disc and the corona. This model features a self-consistent treatment of the thermal, Compton scattering and the reflection component: KERRBB2 describes the thermal component and supplies the seed photons for SIMPLR \citep[a modified version of SIMPL,][]{Steiner2011} to generate the Compton component; while a portion of the Compton component will escape to reach an observer, the remains \citep[refer as SIMPLC,][]{Steiner2011} will strike back to the disc to generate the reflected component. The reflection fraction $R_{\rm ref}$ in IREFLECT \citep{Magdziarz1995}, defined as the ratio of the Compton photons striking back to the disc to that escaping to infinity, is restricted to negative value thereby only the reflected component is returned by IREFLECT. It is linked to the reflection constant parameter $x$ in SIMPLR via the relation $x=1+|R_{\rm ref}|$ \citep{Gou2011}. We set the elemental abundance to unity and the iron abundance {$A_{\rm Fe}$} to five times the solar abundance \citep{Bharali2019, Buisson2019, Xu2020}. The disk temperature $T_{\rm in}$ is fixed at the value returned by DISKBB \citep[M1, refer to][]{Gou2011}. The ionization parameter $\xi$ is fixed at $1000$ \citep[i.e. log$(\xi)=3$,][]{Xu2020, Buisson2019}, as it is difficult to be constrained. Finally we use the KERRCONV \citep{Brenneman2006} to apply relativistic effects assuming an unbroken emissivity profile with index $q=3$. The key parameters in KERRBB2 and KERRCONV are linked together. 

\begin{figure}
    \centering\includegraphics[width=0.48\textwidth]{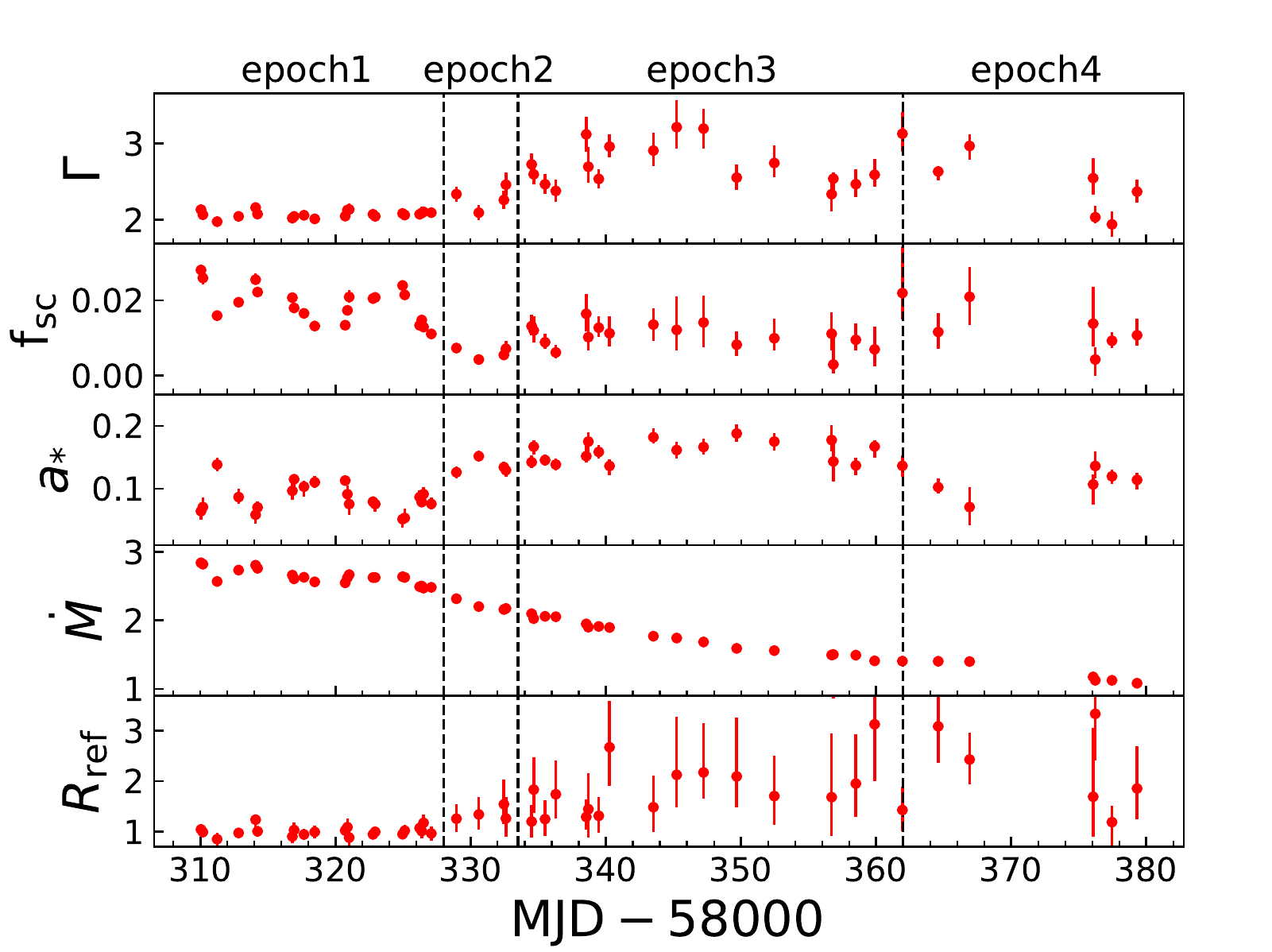}
    \caption{Evolution of the spectral parameters of M4. $\Gamma$ is the photon index; $f_{\rm sc}$ is the scattering fraction; $a_*$ is the spin; $\dot{M}$ is the mass accretion rate in units of $10^{18}$ \,g/s; $R_{\rm ref}$ is the reflection fraction.}
    \label{fig:krrbbReflect}
\end{figure}

\begin{figure}
    \centering\includegraphics[width=0.48\textwidth]{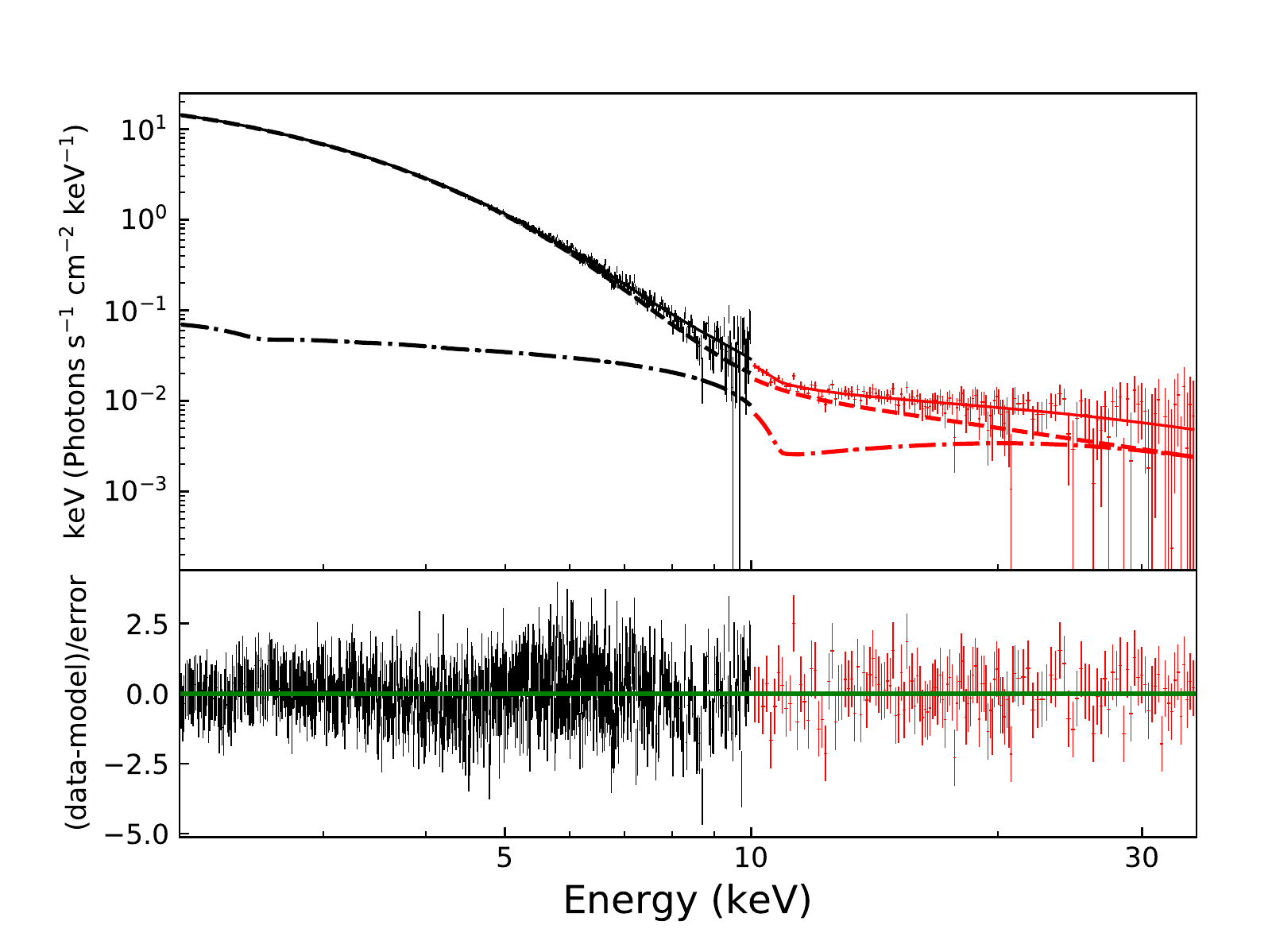}
    \caption{A representative spectrum (ObsID P011466110001) of MAXI J1820 fitted with M4. The dashed line represents the thermal plus the observed Compton components while the dashed-dotted line denotes the reflection component.}
    \label{fig:krrbbReflectSpectrum}
\end{figure}

We show the fitting results in Figure \ref{fig:krrbbReflect} and Table \ref{tbl:M4}. A representative plot of the fitting spectrum is given in Figure \ref{fig:krrbbReflectSpectrum}. The photon index $\Gamma$, the spin, $\dot{M}$ and the non-thermal luminosity are similar to that of M2. It is worth mentioning that the evolution trend of $f_{\rm sc}$ is clearer. More importantly, $R_{\rm ref}$ shows an increasing trend around MJD 58330. The increasing reflection fraction indicates that more Compton photons strike to the disc than escape to infinity. We also have tried to thaw the ionization parameter $\xi$, which yields a similar evolution trend of $R_{\rm ref}$. We also test the fits by fixing the CONSTANT of ME to 1 for all spectral fittings and find that the evolution trends of all parameters are barely affected. Moreover, as IREFLECT*SIMPLC returns a reflected spectrum without emission lines, following \citet{Gou2011}, we have tried to include an extra broad iron line model KERRDISK \citep{Brenneman2006} to evaluate its influence on the spin measurement and found the fitting results are almost unaffected. In addition, note that \citet{Fabian2020} identified another sub-dominate thermal component which was attributed to the additional emission from within the plunge region. We thus have tried to mimic this effect by including an inner disk torque in KERRBB2, which can give additional thermal emission right at the edge of the ISCO, and evaluated its impact on the spin measurement. $\eta=0.1$ is used since MAXI J1820 has a thin accretion disc and even the relatively thick disk would have a small torque \citep[$\eta\sim0.2$,][]{Li2005}. The evolution trend of all fitting parameters are found seldom affected and the value of the spin just decreases by $\sim0.02$ for M2 and $\sim0.04$ for M4. Finally, basing on M4, the best-fit value of $a_{*}$ during the third epoch ranges from 0.13 to 0.19, with the average of 0.16.

\begin{figure*}
    \centering
    \subfigure{
      \includegraphics[width=5.5 cm]{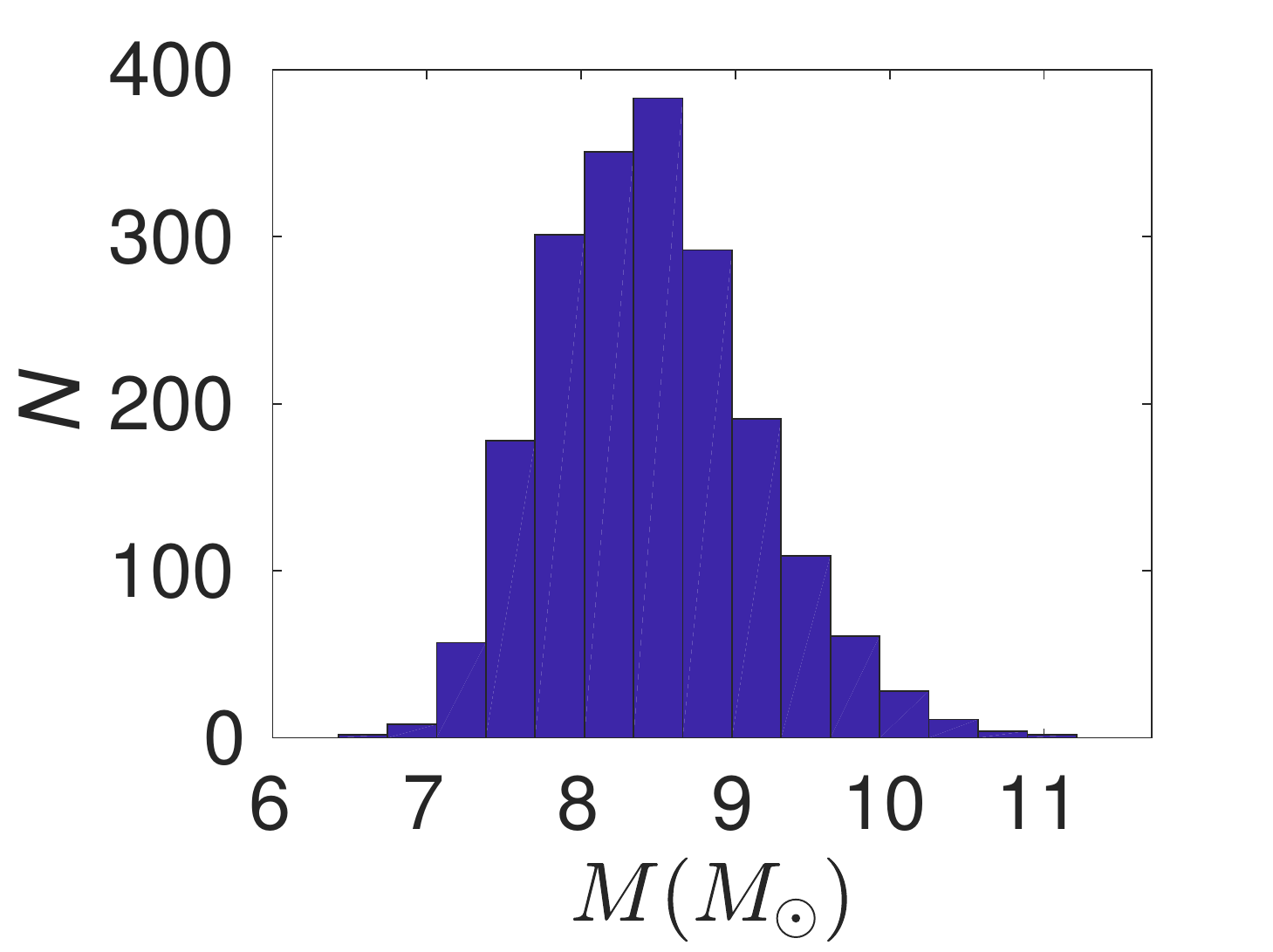}
      \label{fig:1}
    }
    \subfigure{
      \includegraphics[width=5.5 cm]{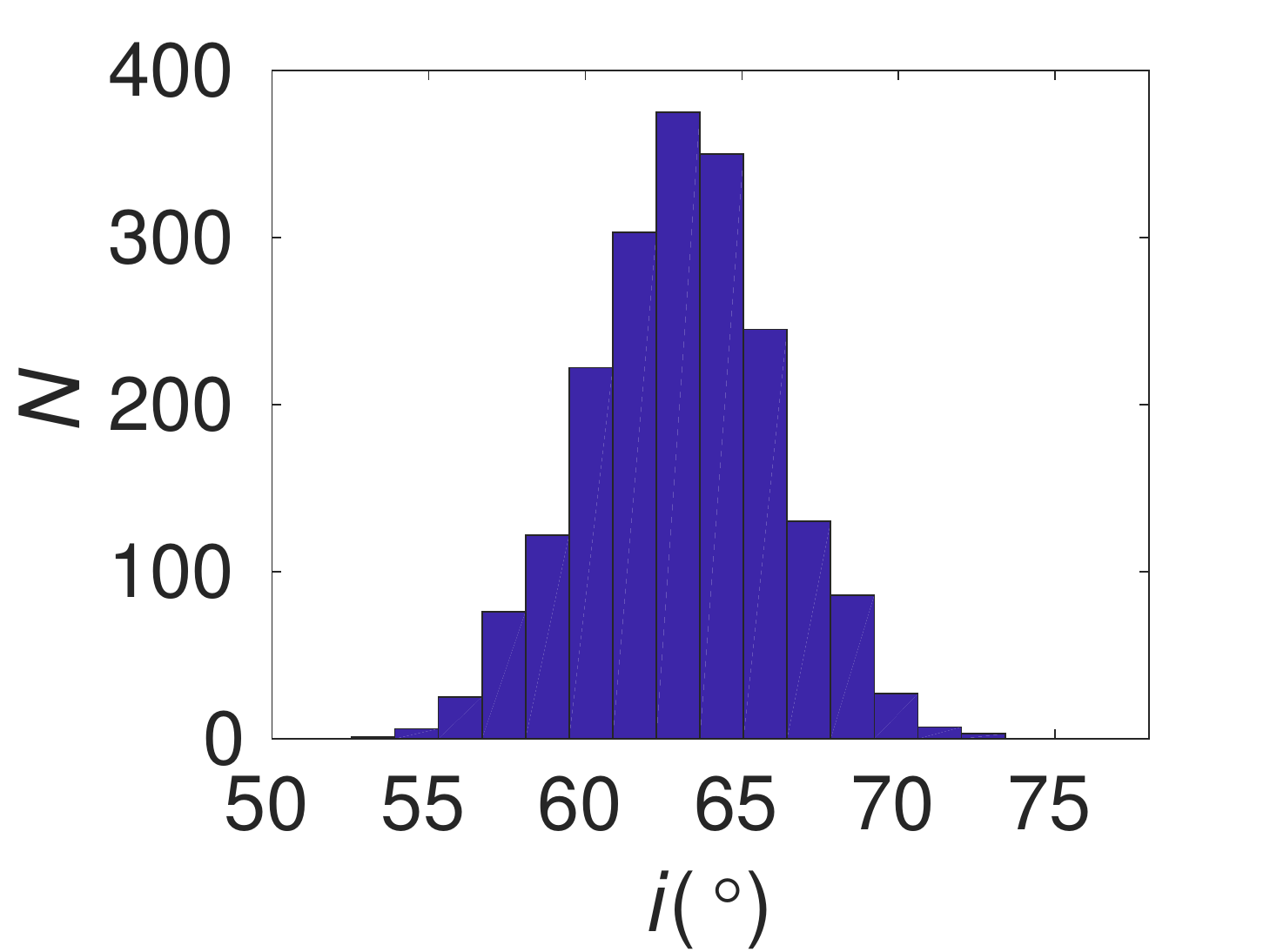}
      \label{fig:2}
    }
      \subfigure{
      \includegraphics[width=5.5 cm]{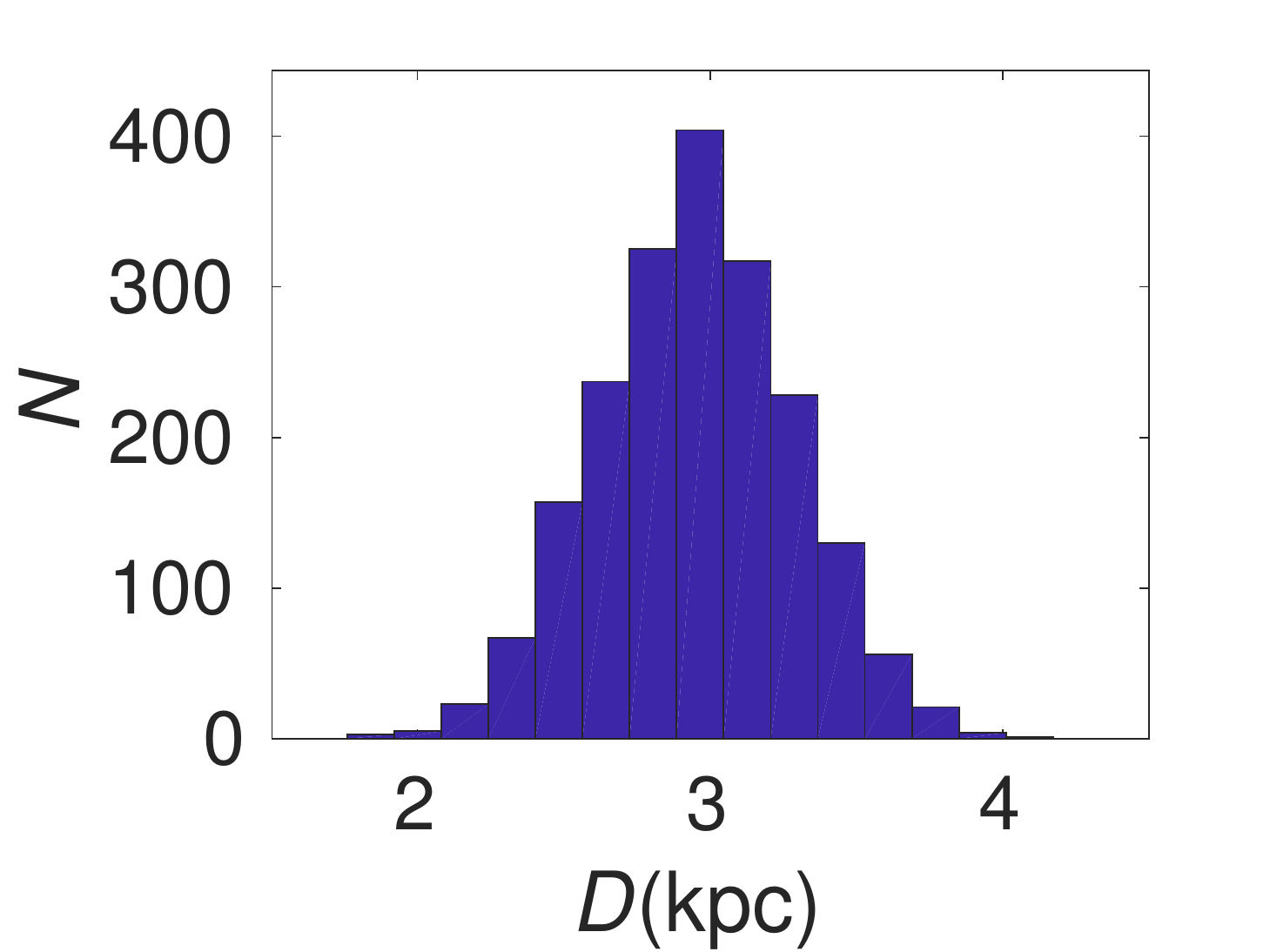}
      \label{fig:3}
    }
\caption{The sampled distributions of the Mass $M$, the inclination $i$ and the distance $D$. Each panel contains 2000 data points.} 
\label{fig:MID}
\end{figure*}

\begin{figure}
    \centering\includegraphics[width=0.48\textwidth]{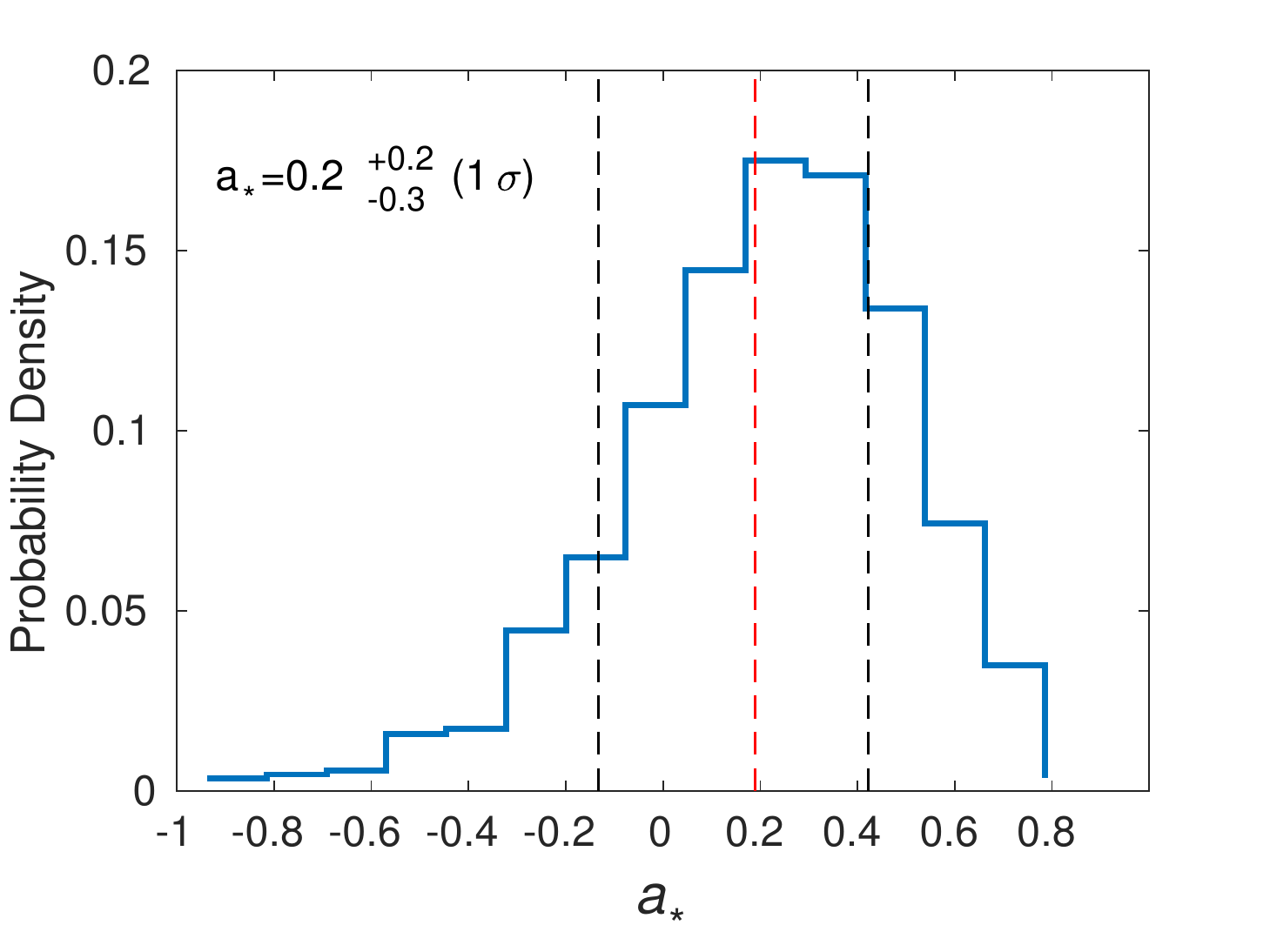}
    \caption{The probability distribution of $a_*$ resulting from the Monte Carlo analysis. The black dashed lines imply the $68.3\%$ (1$\sigma$) error, and the red dashed line represents the best fitting value.} 
    \label{fig:probability}
\end{figure}

\subsection{Error analysis}

\citet{Gou2011} has found that the combined uncertainties of the three key parameters $M$, $i$ and $D$ dominate the error budge in the CF measurements of spin. Following the prescription described in \citet{Gou2011}, we perform Monte Carlo (MC) simulations to analyze the error. Because $M$ and $i$ are not independent, we decouple them by the aid of the mass function $f(M)=M^3sin^3i/(M+M_{\rm opt})^2=5.18\pm0.15\,M_{\odot}$, where $M_{\rm opt}$ \citep[$=0.61^{+0.13}_{-0.12}\,M_{\odot}$ for MAXI J1820,][]{Torres2020} is the mass of the optical companion. Assuming that the value of the mass function, the mass of the companion and the inclination are independent and normally distributed, we can computed the mass of the black hole with these quantities. Thus, the procedure is as follows: for the spectrum in the third epoch (we use ObsID P011466110902 as a representative spectrum), we (1) generate 2000 parameter sets for $f(M)$, $M_{\rm opt}$, $i$, and $D$; (2) solved for $M$ for a set of ($f(M)$, $i$, $M_{opt}$); (3) calculate the look-up table for each set (Figure~\ref{fig:MID}); (4) re-fit the spectrum with M2 to determine $a_*$ and obtain its distribution. The final probability distribution of $a*$ is shown in Figure \ref{fig:probability}. Using the average value from the third epoch of M2, $a_*$ would be $0.2^{+0.2}_{-0.3}$ at the 1$\sigma$ level of confidence.

\section{Discussion}

\label{sec:dis}

In this paper, we report the results of a broad-band spectral characterization of the BH MAXI J1820, as observed with \emph{Insight}-HXMT in 2018 during its HS state. By fitting the observed X-ray spectra with a simple continuum model (M1) consisting of an absorbed multicolour blackbody from the disc (DISKBB) plus a powerlaw component (POWERLAW) from the corona, we find that both $T_{\rm in}$ and the absolute disc luminosity decrease monotonically while $R_{\rm in}$ shows a dramatic evolution, which remains stable till MJD $\sim58330$ then drops to stay at a new stable level and finally increases back (Figure~\ref{fig:diskbb}). By studying the relation between the disc luminosity $L_{\rm disk}$ and the inner disc temperature $T_{\rm in}$, we find that $L_{\rm disk}$ follows $L_{\rm disk}\propto T_{\rm in}^4$ only when $R_{\rm in}$ reaches a minimal size from MJD $\sim58333$ to $\sim58360$ (epoch 3, see Figure~\ref{fig:LT4}). Then we apply a more relativistic model (M2) consisting of KERRBB2 and SIMPL to fit the spectrum and derive the spin. As shown in Figure~\ref{fig:krrbbSimpl}, the evolution of the spin is opposite to that of $R_{\rm in}$, which stays at $\sim0.1$ at the first epoch then increases to $\sim0.18$ at the third epoch then decreases. A drop around MJD $58330$ is observed in the evolution of the non-thermal luminosity, indicating that the hard component is decreasing (Figure~\ref{fig:nonthermal}). We then replace KERRBB2 by its cousin model KERRBB (M3) and find that the spin also shows a similar evolution trend (Figure~\ref{fig:spin}). Finally, we perform a fitting with a more sophisticated model (M4) to evaluate the influence of the excess around 5--10 keV on the spin measurement and find that the derived spin also follows a similar evolution trend (Figure~\ref{fig:krrbbReflect}). Meanwhile, the model provides an increasing reflection fraction ($R_{\rm ref}$) around MJD $58330$, which indicates that more Compton photons strike to the disc than escape to infinity.

\subsection{BH spin}

As presented above, all relativistic disc models (M2-M4) suggest an evolution of the inferred spin during the ``HS state'' defined by \citet{Shidatsu2019}. Physical spin evolution for stellar mass BHs due to accretion is on a timescale of $10^9$ years \citep[e.g.,][]{Chen1997}, thus the inferred spin evolution for MAXI J1820 is apparently nonphysical and must be related to problems in estimating the radius of ISCO. Since $R_{\rm ISCO}$ is assumed to be $R_{\rm in}$ here, $R_{\rm in}$ should also have undergone an evolution, which is consistent with the results of model M1. We indeed observe $R_{\rm in}$ drops down around MJD 58330 and then stays stable at the third epoch. Moreover, $L_{\rm disk}$ also agrees with the expected relation $L_{\rm disk}\propto T_{\rm in}^4$ at this epoch. Although the $L_{\rm disk}\propto T_{\rm in}^4$ dependence would be disturbed when the disk energy is transported to the corona and then, if the corona is very compact, is lost into the black hole, considering $f_{\rm sc}$ is relatively small here, this effect might be negligible. Therefore, we conclude that the disc extends to the ISCO at the third epoch and the spin value ($a_{*}=0.2^{+0.2}_{-0.3}$) obtained in this period is reliable, while other epochs are similar to the intermediate state, in which the disc is still truncated.

The spin measurement is consistent with the expectations of previous works \citep{Atri2020, Buisson2019}, which favor that MAXI J1820 is a slowly spinning system. Our work is also consistent with another independent and similar study \citep{Zhao2020b}, which was posted on arXiv when we were about to submit the current manuscript. They also measured the spin of MAXI J1820 via the CF method using \emph{Insight}-HXMT data and constrained the spin to be $0.13^{+0.07}_{-0.10}$, which is consistent with our results ($a_{*}=0.2^{+0.2}_{-0.3}$). Moreover, a similar evolution trend of the spin can also be found in their Table 3. \citet{Fabian2020} also derived a low spin when adopting the inclination from \citet{Buisson2019}, i.e. $\sim40$\,degrees, but implied a very rapidly spinning retrograde black hole if a high inclination was used. However, almost all the observational results favor a high inclination of MAXI J1820, e.g. the strong absorption dips in X-ray light curves \citep{Homan2018,Kajava2019}, the disk grazing eclipse by the donor star in the H$\alpha$ line \citep{Torres2019,Torres2020} and the jet inclination \citep{Atri2020}. Moreover, \citet{Zdziarski2021} recently reanalyzed the \textit{NuSTAR} data using \citet{Buisson2019} model but with the newly updated \textit{NuSTAR} calibration and the new version of the reflection model RELXILL, and found a high inclination of ${69^{+1}_{-9}}^{\circ}$ as well. In order to address the apparent discrepancy with the \citet{Fabian2020} result, we have tested the fits by using the same \textit{NICER} observation (ObsID 1200120236) and the same disk model (KERRBB) as \citet{Fabian2020}, but fixing the three key parameters at the values adopted in our manuscript (i.e. $M=8.48\,M_{\odot}$, $D=2.96$\,kpc, $i=63^{\circ}$); the spin measurement derived is consistent with the result described in Section \ref{sec:spectral properties} above. Since either reducing $M$ or increasing $D$ would increase $R_{\rm in}/R_{\rm g}$ and thereby decrease the derived spin, it is quite natural to explain the discrepancy as a lower BH mass ($7-8\,M_{\odot}$) and a larger distance (3.5\,kpc) were used in \citet{Fabian2020}, and we could repeat their results if using the same $M$ and $D$. Therefore, we conclude that the spin of MAXI J1820 is low, based on the updated mass, inclination angle and distance.

The low spin value of MAXI J1820 is also supported by the recent studies of QPOs of this source, which found that the highest frequency of the QPO of MAXI J1820 is low \citep[$<1$ Hz,][]{Stiele2020,Ma2020}. According to the model of Lense-Thirring procession, especially the relativistic precession model, the highest precession frequency (or the highest QPO frequency) is approximately proportional to BH spin \citep{Ingram2020}. This expectation has been confirmed by abundant observed evidence, which show that the highest observed QPO frequency does indeed vary from one BH system to the other \citep{Motta2014, Ingram2014, Franchini2017}.

Moreover, the low spin of MAXI J1820 may challenge the widely held concept that powerful jets are driven by BH spins \citep[i.e. the BZ mechanism,][]{BZ1977} and instead supports the BP mechanism \citep{BP1982}, in which the jet is powered by extracting the energy of the magnetized accretion disk. When the BH spin is less than 0.4, the BP mechanism is more efficient to provide energy to jet than the BZ mechanism \citep{Steiner2011}. For MAXI J1820, given the low spin ($\sim 0.2$), its strong radio flare, corresponding to the launch of superluminal ejecta \citep{Homan2020}, may be mainly driven by the accretion disc rather than by the BH hole spin. 

\subsection{Evolution of the accretion flows}

The evolution of the spin and $R_{\rm in}$ suggest that the inner disc has changed, while the evolution of other parameters, i.e., hardness ratio, non-thermal luminosity and reflection fraction, can help us to understand the full picture of the accretion flows. 

\begin{figure}
    \centering\includegraphics[width=0.48\textwidth]{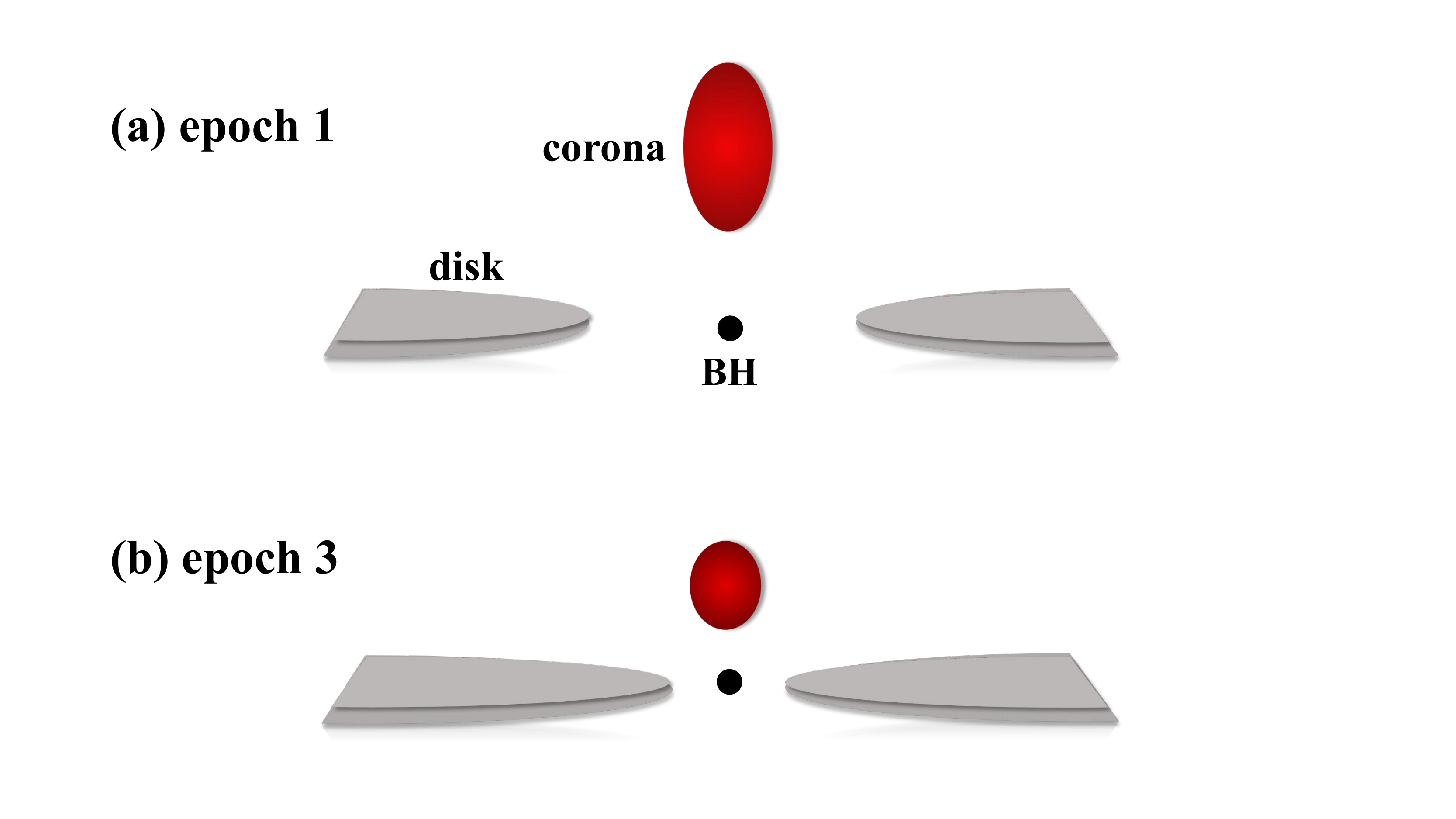}
    \caption{Schematic of the coronal collapsing model. Due to the condensation via inverse Compton scattering, the corona may collapse into a part of the inner disc.}
    \label{fig:G1}
\end{figure}

When the spin increases and $R_{\rm in}$ decreases at MJD $\sim58330$, we observe a decline of hardness ratio as well as the non-thermal luminosity, and an increase of the reflection fraction. These results indicate a dynamical interaction between the inner disc and corona, and could be self-consistently interpreted in the context of the disc accretion fed by condensation of the hot corona (Figure~\ref{fig:G1}) \citep{Liu2006, Meyer2007, Qiao2012}. In this mechanism, condensation via inverse Compton scattering would lead to the collapse of the corona and thus the inward extension of the inner disc edge \citep{Liu2013}, which coincides with the decline of both the non-thermal flux and the hardness ratio reported in our work. Moreover, the contraction in the corona's spatial extent and the extension of the disc will increase the opening angle of the corona to the disc and thus increase the reflection fraction. The corona's contraction has already been observed in the hard state of MAXI J1820 with NICER observations \citep{Kara2019}. 

\begin{figure}
    \centering\includegraphics[width=0.48\textwidth]{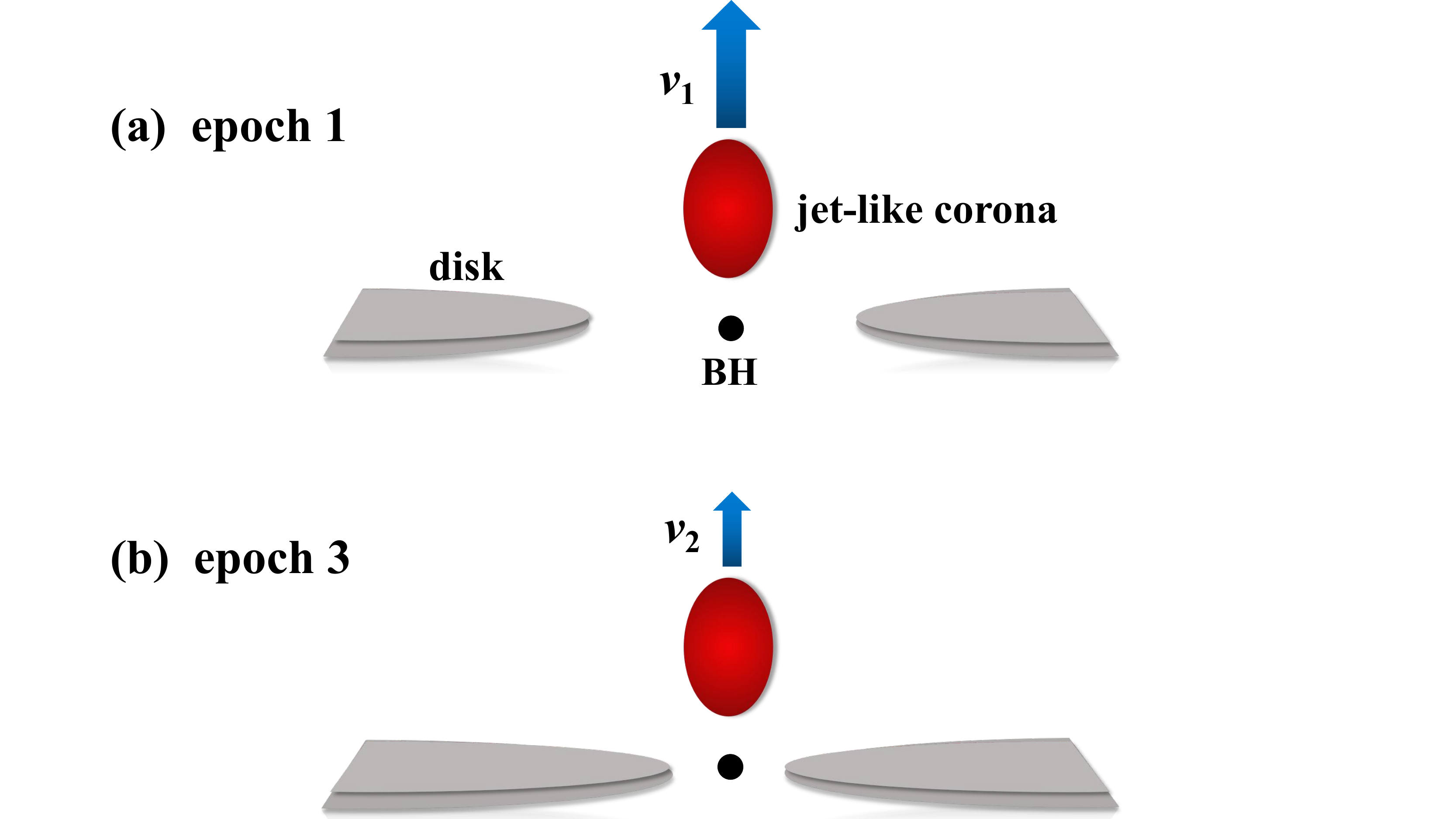}
    \caption{Schematic of the jet-like corona decelerating model. The bulk velocity of epoch 3 is lower than epoch 1 ($v2<v1$).} 
    \label{fig:G2}
\end{figure}

Motivated by the work of \citet{You2020} for MAXI J1820 and the detection of the relativistic ejecta in both radio \citep{Bright2020} and X-ray \citep{Espinasse2020} bands in the soft state, there is another scenario (see Figure~\ref{fig:G2}). That is, the hot electrons in the corona have outward bulk motion, acting like the jet base. Even if the jet-like corona does not collapse, if its bulk velocity decreases, because of the beaming effect, the hard flux as well as the hardness ratio will decrease, and the reflection fraction will increase as more photons can illuminate the accretion disk \citep{Markoff2005}. In this process, the inner disc radius extends to the ISCO. The disk extension should be related to the jet-like corona deceleration through some inherent effects. However, the mechanism of the deceleration of the jet-like corona \citep[e.g.,][]{ChenL2020} and how the deceleration is related to the observed disc extension are still unclear, further theoretical understanding is needed.

\section{Conclusion}

\label{sec:conclusion}

We have carefully studied the spectral evolution of MAXI J1820 during its HS state, which shows obvious evolution of the fitting parameters, in particular of the inferred spin with the CF method. By studying the inner disk evolution and the relation between the disc luminosity $L_{\rm disk}$ and the inner disc temperature $T_{\rm in}$, we associate it with the inexact estimation of the ISCO and find that only at the third epoch did the inner disc reach the ISCO and the spin value is valid.  While the evolution of the spin and $R_{\rm in}$ suggest that the inner disc has changed, the evolution of other parameters, i.e., hardness ratio, non-thermal luminosity and reflection fraction, can be interpreted as the collapse in the corona caused by the condensation mechanism, or be interpreted as the deceleration of a jet-like corona. Investigation of the changes in the accretion flow geometry thus offers some useful clues for understanding the accretion physics in the vicinity of the BH.

The spin of MAXI J1820 is $a_*=0.2^{+0.2}_{-0.3}$ based on the measurement of the third epoch when the inner disc extends to the ISCO. The low spin is also supported by previous spectral and timing results. Such a slowly spinning BH harboring a powerful jet indicates that BZ mechanism is not primary in driving its powerful jets but the BP effect is the main cause.

\section*{Acknowledgements}
We thank the referee (C. Reynolds) for helpful comments and suggestions. We appreciate useful discussions with B. You and L. Chen. This work made use of data from the \emph{Insight}-HXMT mission, a project funded by China National Space Administration (CNSA) and the Chinese Academy of Sciences (CAS). This work is supported by the National Key R\&D Program of China (2016YFA0400800, 2016YFA0400803) and the National Natural Science Foundation of China under grants U2038102, U1838115, U1838201, 11673023, U1938104, U1838111, 11473027, U1838202, 11733009, U1838104, and U1938101.

\section{DATA AVAILABILITY}
The data that support the findings of this study are available from \emph{Insight}-HXMT’s data archive (http://enghxmt.ihep.ac.cn).

\linespread{1.8}
\begin{table*}
    \caption{\emph{Insight}-HXMT Observations of MAXI J1820 during the HS state.}
    \label{tbl:Observations}
    \begin{center}
    \begin{tabular}{ccccccc}
    \hline\hline
    ObsID & Observed date & Observed start date & LE Exposure time & LE Counts rate & ME Exposure time & ME Counts rate \\ 
    & (MJD) & & (s) & (cts/s) & (s) & (cts/s) \\ \hline
P011466108801 & 58310.04 & 2018-07-11T00:54:27 & 1616 & $1124.3 \pm 0.8  $ & 1800 & $ 39.9  \pm 0.3  $ \\  
P011466108802 & 58310.18 & 2018-07-11T04:17:01 &  898 & $1125.9 \pm 1.1  $ & 1260 & $ 38.6  \pm 0.3  $ \\  
P011466108901 & 58311.23 & 2018-07-12T05:32:24 &  931 & $1062.0 \pm 1.1  $ & 3180 & $ 24.88 \pm 0.17 $ \\  
P011466109001 & 58312.82 & 2018-07-13T19:43:03 & 2135 & $1082.7 \pm 0.7  $ & 3930 & $ 28.53 \pm 0.15 $ \\  
P011466109101 & 58314.08 & 2018-07-15T01:56:30 & 1002 & $1095.5 \pm 1.1  $ & 1410 & $ 35.6  \pm 0.3  $ \\  
P011466109102 & 58314.23 & 2018-07-15T05:23:05 & 1194 & $1083.3 \pm 1.0  $ & 2700 & $ 32.68 \pm 0.19 $ \\  
P011466109201 & 58316.80 & 2018-07-17T19:10:07 & 1197 & $1057.2 \pm 0.9  $ & 2550 & $ 29.52 \pm 0.19 $ \\  
P011466109202 & 58316.93 & 2018-07-17T22:15:11 & 1306 & $1052.1 \pm 0.9  $ & 1710 & $ 27.8  \pm 0.3  $ \\  
P011466109301 & 58317.66 & 2018-07-18T15:51:10 & 1077 & $1042.1 \pm 1.0  $ & 4590 & $ 23.27 \pm 0.14 $ \\  
P011466109401 & 58318.46 & 2018-07-19T10:56:52 & 1257 & $1012.9 \pm 0.9  $ & 4260 & $ 18.41 \pm 0.13 $ \\  
P011466109501 & 58320.71 & 2018-07-21T17:03:42 & 3039 & $1010.6 \pm 0.6  $ & 3780 & $ 18.50 \pm 0.14 $ \\  
P011466109502 & 58320.88 & 2018-07-21T20:59:14 & 1117 & $1028.1 \pm 1.0  $ & 1680 & $ 23.2  \pm 0.3  $ \\  
P011466109503 & 58321.01 & 2018-07-22T00:10:06 &  479 & $1034.3 \pm 1.5  $ & 1020 & $ 26.1  \pm 0.3  $ \\  
P011466109601 & 58322.77 & 2018-07-23T18:25:25 & 2238 & $1017.8 \pm 0.7  $ & 3360 & $ 28.20 \pm 0.18 $ \\  
P011466109602 & 58322.93 & 2018-07-23T22:17:50 & 1373 & $1016.6 \pm 0.9  $ & 2400 & $ 29.8  \pm 0.2  $ \\  
P011466109701 & 58324.96 & 2018-07-25T23:00:04 & 1562 & $991.8  \pm 0.8  $ & 3030 & $ 29.58 \pm 0.18 $ \\ 
P011466109702 & 58325.12 & 2018-07-26T02:47:29 & 1584 & $983.5  \pm 0.8  $ & 2400 & $ 28.2  \pm 0.2  $ \\ 
P011466109801 & 58326.22 & 2018-07-27T05:17:00 & 2575 & $943.8  \pm 0.6  $ & 3180 & $ 17.13 \pm 0.15 $ \\ 
P011466109802 & 58326.38 & 2018-07-27T09:01:17 & 2247 & $941.7  \pm 0.7  $ & 3000 & $ 17.79 \pm 0.16 $ \\ 
P011466109803 & 58326.51 & 2018-07-27T12:12:12 & 2213 & $939.6  \pm 0.7  $ & 2670 & $ 15.66 \pm 0.18 $ \\ 
P011466109901 & 58327.08 & 2018-07-28T02:00:03 & 2543 & $919.7  \pm 0.6  $ & 5220 & $ 12.28 \pm 0.12 $ \\ 
P011466110001 & 58328.94 & 2018-07-29T22:35:48 & 2704 & $880.9  \pm 0.6  $ & 5430 & $ 5.19  \pm 0.11 $ \\ 
P011466110101 & 58330.60 & 2018-07-31T14:22:58 & 2262 & $850.8  \pm 0.6  $ & 5640 & $ 4.97  \pm 0.11 $ \\ 
P011466110301 & 58332.46 & 2018-08-02T10:55:26 & 2104 & $807.8  \pm 0.6  $ & 5730 & $ 3.89  \pm 0.11 $ \\ 
P011466110302 & 58332.61 & 2018-08-02T14:35:40 & 2571 & $809.0  \pm 0.6  $ & 4980 & $ 3.87  \pm 0.12 $ \\ 
P011466110401 & 58334.51 & 2018-08-04T12:13:41 & 2831 & $788.0  \pm 0.5  $ & 6420 & $ 3.85  \pm 0.10 $ \\ 
P011466110402 & 58334.66 & 2018-08-04T15:55:25 & 1538 & $781.9  \pm 0.7  $ & 2670 & $ 5.27  \pm 0.16 $ \\ 
P011466110701 & 58335.50 & 2018-08-05T12:05:00 & 4219 & $764.3  \pm 0.4  $ & 7110 & $ 3.34  \pm 0.09 $ \\ 
P011466110801 & 58336.30 & 2018-08-06T07:10:03 & 2170 & $756.1  \pm 0.6  $ & 5220 & $ 3.39  \pm 0.11 $ \\ 
P011466110901 & 58338.55 & 2018-08-08T13:14:17 & 3284 & $722.3  \pm 0.5  $ & 5850 & $ 2.85  \pm 0.10 $ \\ 
P011466110902 & 58338.71 & 2018-08-08T16:58:34 &  898 & $722.4  \pm 0.9  $ & 2310 & $ 3.60  \pm 0.16 $ \\ 
P011466111001 & 58339.48 & 2018-08-09T11:30:11 & 3043 & $712.6  \pm 0.5  $ & 5970 & $ 4.15  \pm 0.10 $ \\ 
P011466111101 & 58340.28 & 2018-08-10T06:35:15 & 2176 & $688.0  \pm 0.6  $ & 5310 & $ 2.69  \pm 0.11 $ \\ 
P011466111301 & 58343.52 & 2018-08-13T12:31:11 & 2155 & $656.7  \pm 0.6  $ & 4320 & $ 2.86  \pm 0.12 $ \\ 
P011466111401 & 58345.25 & 2018-08-15T05:52:31 & 2692 & $623.0  \pm 0.5  $ & 4800 & $ 1.70  \pm 0.11 $ \\ 
P011466111501 & 58347.23 & 2018-08-17T05:35:56 & 2514 & $601.5  \pm 0.5  $ & 4800 & $ 1.92  \pm 0.11 $ \\ 
P011466111601 & 58349.69 & 2018-08-19T16:27:55 & 1735 & $568.3  \pm 0.6  $ & 3660 & $ 2.93  \pm 0.12 $ \\ 
P011466111701 & 58352.47 & 2018-08-22T11:17:57 & 1647 & $535.4  \pm 0.6  $ & 3750 & $ 2.64  \pm 0.11 $ \\ 
P011466111801 & 58356.72 & 2018-08-26T17:09:14 &  599 & $511.9  \pm 0.9  $ &  960 & $ 4.3   \pm 0.2  $ \\ 
P011466111802 & 58356.84 & 2018-08-26T20:09:34 &  180 & $500.5  \pm 1.7  $ &  900 & $ 4.4   \pm 0.3  $ \\ 
P011466111901 & 58358.51 & 2018-08-28T12:07:17 & 1542 & $478.0  \pm 0.6  $ & 4110 & $ 2.76  \pm 0.11 $ \\ 
P011466112001 & 58359.90 & 2018-08-29T21:32:12 & 1317 & $454.9  \pm 0.6  $ & 2640 & $ 2.30  \pm 0.14 $ \\ 
P011466112101 & 58361.95 & 2018-08-31T22:51:33 & 1594 & $430.6  \pm 0.5  $ & 4650 & $ 2.19  \pm 0.11 $ \\ 
P011466112201 & 58364.60 & 2018-09-03T14:29:34 &  958 & $411.3  \pm 0.7  $ & 2670 & $ 3.42  \pm 0.15 $ \\ 
P011466112301 & 58366.92 & 2018-09-05T22:10:06 &  405 & $393.5  \pm 1.0  $ & 2880 & $ 2.44  \pm 0.14 $ \\ 
P011466112401 & 58376.07 & 2018-09-15T01:42:32 & 1833 & $304.3  \pm 0.4  $ & 3000 & $ 2.40  \pm 0.14 $ \\ 
P011466112402 & 58376.23 & 2018-09-15T05:26:46 &  539 & $301.4  \pm 0.8  $ & 1290 & $ 3.6   \pm 0.2  $ \\ 
P011466112501 & 58377.46 & 2018-09-16T11:07:41 & 4309 & $290.4  \pm 0.3  $ & 3690 & $ 3.09  \pm 0.12 $ \\ 
P011466112601 & 58379.32 & 2018-09-18T07:41:38 & 2775 & $272.5  \pm 0.3  $ & 3870 & $ 2.93  \pm 0.12 $
 \\ \hline
    \end{tabular}%
    \end{center}
\end{table*}

\linespread{1.8}
\begin{table*}
    \caption{Various models used for fitting the \emph{Insight}-HXMT spectra of MAXI J1820, and the corresponding applied energy bands.}
    \label{tbl:models}
    \begin{center}
    \begin{tabular}{ccc}
    \hline\hline
    Model & Energy bands  \\ \hline
    M1:\,CONSTANT*TBABS*(DISKBB+POWERLAW) & 2-5\,keV\,(LE),\,10-20\,keV\,(ME) \\ 
    M2:\,CONSTANT*TBABS*(SIMPL*KERRBB2) & 2-5\,keV\,(LE),\,10-20\,keV\,(ME) \\
    M3:\,CONSTANT*TBABS*(SIMPL*KERRBB) & 2-5\,keV\,(LE),\,10-20\,keV\,(ME) \\
    M4:\,CONSTANT*TBABS*(SIMPLR*KERRBB2+KERRCONV*(IREFLECT*SIMPLC)) & 2-10\,keV\,(LE),\,10-35\,keV\,(ME) \\ \hline
    \end{tabular}%
    \end{center}
\end{table*}

\linespread{1.8}
\begin{table*}
    \caption{Fitting results for Model M1. $T_{\rm in}$ is the inner disc temperature; $R_{\rm in}$ is the apparent inner disc radius in units of $R_{\rm g}$, where $R_{\rm g}=GM/c^2=12.5$\,km for $M=8.48\,M_{\odot}$; $\Gamma$ is the photon index; $N_{\rm pl}$ is the normalization of POWERLAW.}
    \label{tbl:M1}
    \begin{center}
    \begin{tabular}{ccccccc}
    \hline\hline
    ObsID & $T_{\rm in}$ & $R_{\rm in}$ & $\Gamma$ & $N_{\rm pl}$ & $\chi^{2} / \rm dof$ & Reduced-$\chi^{2}$ \\ 
    & (keV) & $(R_{\rm g})$ & & & & \\\hline
P011466108801 & $0.749  _{-0.002} ^{+0.002} $ & $4.47 _{-0.03} ^{+0.03}$ & $2.12 _{-0.06} ^{+0.06}$ & $1.9   _{- 0.3 } ^{+0.3} $ & 368.7/433 & 0.85 \\
P011466108802 & $0.750  _{-0.003} ^{+0.003} $ & $4.47 _{-0.04} ^{+0.04}$ & $1.98 _{-0.08} ^{+0.08}$ & $1.3   _{- 0.2 } ^{+0.3} $ & 355.9/433 & 0.82 \\
P011466108901 & $0.753  _{-0.002} ^{+0.002} $ & $4.35 _{-0.03} ^{+0.03}$ & $1.99 _{-0.07} ^{+0.07}$ & $0.85  _{- 0.13} ^{+0.16}$ & 353.0/433 & 0.82 \\
P011466109001 & $0.7511 _{-0.0017}^{+0.0018}$ & $4.43 _{-0.03} ^{+0.03}$ & $1.99 _{-0.05} ^{+0.05}$ & $0.97  _{- 0.12} ^{+0.15}$ & 372.1/433 & 0.86 \\
P011466109101 & $0.743  _{-0.003} ^{+0.003} $ & $4.50 _{-0.03} ^{+0.03}$ & $2.24 _{-0.08} ^{+0.08}$ & $2.3   _{- 0.4 } ^{+0.5} $ & 365.5/433 & 0.84 \\
P011466109102 & $0.746  _{-0.002} ^{+0.002} $ & $4.46 _{-0.03} ^{+0.03}$ & $2.10 _{-0.06} ^{+0.06}$ & $1.5   _{- 0.2 } ^{+0.2} $ & 383.2/434 & 0.88 \\
P011466109201 & $0.746  _{-0.002} ^{+0.002} $ & $4.43 _{-0.03} ^{+0.03}$ & $2.02 _{-0.06} ^{+0.06}$ & $1.20  _{- 0.17} ^{+0.2} $ & 368.3/433 & 0.85 \\
P011466109202 & $0.747  _{-0.002} ^{+0.002} $ & $4.42 _{-0.03} ^{+0.03}$ & $2.05 _{-0.09} ^{+0.09}$ & $1.1   _{- 0.2 } ^{+0.3} $ & 383.7/434 & 0.88 \\
P011466109301 & $0.744  _{-0.002} ^{+0.002} $ & $4.46 _{-0.03} ^{+0.03}$ & $2.05 _{-0.06} ^{+0.06}$ & $0.93  _{- 0.13} ^{+0.15}$ & 404.7/433 & 0.93 \\
P011466109401 & $0.7444 _{-0.0019}^{+0.0019}$ & $4.41 _{-0.03} ^{+0.03}$ & $2.01 _{-0.07} ^{+0.07}$ & $0.68  _{- 0.11} ^{+0.14}$ & 389.1/433 & 0.90 \\
P011466109501 & $0.7437 _{-0.0016}^{+0.0016}$ & $4.41 _{-0.03} ^{+0.02}$ & $2.06 _{-0.08} ^{+0.08}$ & $0.77  _{- 0.14} ^{+0.17}$ & 313.8/433 & 0.72 \\
P011466109502 & $0.740  _{-0.002} ^{+0.002} $ & $4.46 _{-0.03} ^{+0.03}$ & $2.22 _{-0.11} ^{+0.11}$ & $1.4   _{- 0.3 } ^{+0.4} $ & 309.9/433 & 0.72 \\
P011466109503 & $0.738  _{-0.003} ^{+0.003} $ & $4.50 _{-0.04} ^{+0.04}$ & $2.18 _{-0.11} ^{+0.11}$ & $1.5   _{- 0.4 } ^{+0.5} $ & 381.0/433 & 0.88 \\
P011466109601 & $0.7387 _{-0.0018}^{+0.0019}$ & $4.46 _{-0.03} ^{+0.03}$ & $2.06 _{-0.06} ^{+0.06}$ & $1.17  _{- 0.17} ^{+0.2} $ & 385.5/433 & 0.89 \\
P011466109602 & $0.736  _{-0.002} ^{+0.002} $ & $4.48 _{-0.03} ^{+0.03}$ & $2.08 _{-0.07} ^{+0.07}$ & $1.3   _{- 0.2 } ^{+0.3} $ & 383.4/434 & 0.88 \\
P011466109701 & $0.732  _{-0.002} ^{+0.002} $ & $4.50 _{-0.03} ^{+0.03}$ & $2.09 _{-0.06} ^{+0.06}$ & $1.3   _{- 0.2 } ^{+0.2} $ & 377.0/433 & 0.87 \\
P011466109702 & $0.732  _{-0.002} ^{+0.002} $ & $4.50 _{-0.03} ^{+0.03}$ & $2.05 _{-0.07} ^{+0.07}$ & $1.12  _{- 0.18} ^{+0.2} $ & 431.4/433 & 0.99 \\
P011466109801 & $0.7310 _{-0.0018}^{+0.0018}$ & $4.46 _{-0.03} ^{+0.03}$ & $2.12 _{-0.09} ^{+0.09}$ & $0.84  _{- 0.17} ^{+0.2} $ & 339.5/433 & 0.78 \\
P011466109802 & $0.7298 _{-0.0018}^{+0.0019}$ & $4.47 _{-0.03} ^{+0.03}$ & $2.16 _{-0.09} ^{+0.09}$ & $0.97  _{- 0.19} ^{+0.2} $ & 383.9/433 & 0.89 \\
P011466109803 & $0.7313 _{-0.0019}^{+0.0019}$ & $4.44 _{-0.03} ^{+0.03}$ & $2.21 _{-0.11} ^{+0.11}$ & $1.0   _{- 0.2 } ^{+0.3} $ & 352.6/433 & 0.81 \\
P011466109901 & $0.7288 _{-0.0016}^{+0.0016}$ & $4.47 _{-0.03} ^{+0.03}$ & $2.10 _{-0.09} ^{+0.09}$ & $0.57  _{- 0.12} ^{+0.15}$ & 352.8/433 & 0.81 \\
P011466110001 & $0.7269 _{-0.0015}^{+0.0016}$ & $4.39 _{-0.04} ^{+0.03}$ & $2.9  _{-0.3 } ^{+0.3 }$ & $1.8   _{- 0.8 } ^{+1.8} $ & 323.3/433 & 0.75 \\
P011466110101 & $0.7262 _{-0.0017}^{+0.0017}$ & $4.36 _{-0.03} ^{+0.03}$ & $2.7  _{-0.3 } ^{+0.3 }$ & $1.0   _{- 0.5 } ^{+1.0} $ & 312.8/433 & 0.72 \\
P011466110301 & $0.7192 _{-0.0015}^{+0.0015}$ & $4.36 _{-0.08} ^{+0.03}$ & $3.1  _{-0.4 } ^{+0.4 }$ & $2.0   _{- 1.3 } ^{+3}   $ & 354.9/433 & 0.82 \\
P011466110302 & $0.7183 _{-0.0018}^{+0.002} $ & $4.30 _{-0.07} ^{+0.05}$ & $3.43 _{-0.2 } ^{+0.18}$ & $5     _{- 3   } ^{+3}   $ & 303.5/433 & 0.70 \\
P011466110401 & $0.723  _{-0.002} ^{+0.002} $ & $4.13 _{-0.07} ^{+0.06}$ & $3.55 _{-0.15} ^{+0.12}$ & $7     _{- 2   } ^{+3}   $ & 312.3/433 & 0.72 \\
P011466110402 & $0.725  _{-0.003} ^{+0.003} $ & $4.04 _{-0.08} ^{+0.07}$ & $3.53 _{-0.14} ^{+0.12}$ & $9     _{- 3   } ^{+3}   $ & 283.4/433 & 0.65 \\
P011466110701 & $0.7179 _{-0.0019}^{+0.002} $ & $4.16 _{-0.07} ^{+0.03}$ & $3.61 _{-0.18} ^{+0.14}$ & $7     _{- 2   } ^{+3}   $ & 300.3/433 & 0.69 \\
P011466110801 & $0.712  _{-0.002} ^{+0.003} $ & $4.26 _{-0.08} ^{+0.07}$ & $3.53 _{-0.3 } ^{+0.18}$ & $6     _{- 3   } ^{+3}   $ & 339.7/433 & 0.78 \\
P011466110901 & $0.714  _{-0.002} ^{+0.002} $ & $4.05 _{-0.07} ^{+0.06}$ & $3.75 _{-0.12} ^{+0.10}$ & $10    _{- 2   } ^{+3}   $ & 266.6/433 & 0.62 \\
P011466110902 & $0.714  _{-0.003} ^{+0.004} $ & $4.05 _{-0.10} ^{+0.09}$ & $3.62 _{-0.20} ^{+0.15}$ & $8     _{- 3   } ^{+4}   $ & 361.4/433 & 0.83 \\
P011466111001 & $0.711  _{-0.002} ^{+0.002} $ & $4.08 _{-0.06} ^{+0.06}$ & $3.51 _{-0.15} ^{+0.13}$ & $7     _{- 2   } ^{+2}   $ & 275.5/434 & 0.63 \\
P011466111101 & $0.707  _{-0.002} ^{+0.003} $ & $4.09 _{-0.07} ^{+0.07}$ & $3.64 _{-0.17} ^{+0.14}$ & $7     _{- 2   } ^{+3}   $ & 277.9/433 & 0.64 \\
P011466111301 & $0.705  _{-0.003} ^{+0.003} $ & $3.99 _{-0.07} ^{+0.07}$ & $3.75 _{-0.13} ^{+0.11}$ & $9     _{- 3   } ^{+3}   $ & 284.3/433 & 0.66 \\
P011466111401 & $0.699  _{-0.003} ^{+0.003} $ & $3.99 _{-0.07} ^{+0.06}$ & $3.99 _{-0.12} ^{+0.11}$ & $11    _{- 3   } ^{+3}   $ & 293.1/433 & 0.68 \\
P011466111501 & $0.696  _{-0.003} ^{+0.003} $ & $3.95 _{-0.07} ^{+0.07}$ & $3.92 _{-0.13} ^{+0.11}$ & $10    _{- 3   } ^{+3}   $ & 297.1/433 & 0.69 \\
P011466111601 & $0.685  _{-0.003} ^{+0.003} $ & $3.97 _{-0.08} ^{+0.04}$ & $3.79 _{-0.13} ^{+0.11}$ & $10    _{- 2   } ^{+3}   $ & 354.4/433 & 0.82 \\
P011466111701 & $0.6721 _{-0.0018}^{+0.002} $ & $4.20 _{-0.07} ^{+0.06}$ & $3.41 _{-0.2 } ^{+0.19}$ & $4     _{- 1.7 } ^{+2}   $ & 340.3/433 & 0.79 \\
P011466111801 & $0.667  _{-0.003} ^{+0.004} $ & $4.09 _{-0.12} ^{+0.10}$ & $3.42 _{-0.3 } ^{+0.19}$ & $6     _{- 3   } ^{+3}   $ & 401.1/434 & 0.92 \\
P011466111802 & $0.665  _{-0.006} ^{+0.006} $ & $4.24 _{-0.08} ^{+0.08}$ & $2.6  _{-0.7 } ^{+0.7 }$ & $0.7   _{- 0.6 } ^{+3}   $ & 319.6/365 & 0.88 \\
P011466111901 & $0.6554 _{-0.0019}^{+0.002} $ & $4.22 _{-0.07} ^{+0.06}$ & $3.44 _{-0.2 } ^{+0.18}$ & $3.9   _{- 1.7 } ^{+2}   $ & 295.3/433 & 0.68 \\
P011466112001 & $0.654  _{-0.003} ^{+0.003} $ & $4.08 _{-0.09} ^{+0.08}$ & $3.76 _{-0.18} ^{+0.15}$ & $7     _{- 2   } ^{+3}   $ & 369.4/433 & 0.85 \\
P011466112101 & $0.646  _{-0.003} ^{+0.003} $ & $4.08 _{-0.09} ^{+0.08}$ & $3.76 _{-0.14} ^{+0.12}$ & $8     _{- 2   } ^{+2}   $ & 328.4/433 & 0.76 \\
P011466112201 & $0.633  _{-0.003} ^{+0.003} $ & $4.22 _{-0.10} ^{+0.08}$ & $3.53 _{-0.18} ^{+0.15}$ & $6     _{- 2   } ^{+2}   $ & 409.1/433 & 0.94 \\
P011466112301 & $0.627  _{-0.003} ^{+0.004} $ & $4.30 _{-0.14} ^{+0.10}$ & $3.4  _{-0.4 } ^{+0.4 }$ & $4     _{- 2   } ^{+3}   $ & 365.3/398 & 0.92 \\
P011466112401 & $0.593  _{-0.003} ^{+0.004} $ & $4.35 _{-0.09} ^{+0.08}$ & $3.68 _{-0.13} ^{+0.12}$ & $6.4   _{- 1.7 } ^{+1.9} $ & 399.8/433 & 0.92 \\
P011466112402 & $0.591  _{-0.004} ^{+0.005} $ & $4.51 _{-0.09} ^{+0.08}$ & $3.2  _{-0.4 } ^{+0.4 }$ & $2.3   _{- 1.4 } ^{+2}   $ & 338.4/391 & 0.87 \\
P011466112501 & $0.587  _{-0.002} ^{+0.002} $ & $4.48 _{-0.05} ^{+0.04}$ & $3.30 _{-0.15} ^{+0.13}$ & $3.0   _{- 0.9 } ^{+1.0} $ & 345.7/433 & 0.80 \\
P011466112601 & $0.578  _{-0.002} ^{+0.003} $ & $4.52 _{-0.05} ^{+0.04}$ & $3.31 _{-0.14} ^{+0.12}$ & $3.1   _{- 0.8 } ^{+1.0 }$ & 375.6/433 & 0.87
     \\ \hline
    \footnotetext{Second footnote}    
    \end{tabular}%
    \end{center}
\end{table*}

\linespread{1.8}
\begin{table*}
\caption{Fitting results for Model M2. $\Gamma$ is the photon index; $f_{\rm sc}$ is the scattering fraction; $a_*$ is the spin; $\dot{M}$ is the mass accretion rate in units of $10^{18}$~g/s; $f$ is the spectral hardening factor; $l$ is the Eddington-scaled luminosity.}
\label{tbl:M2}
\begin{center}
\begin{tabular}{ccccccccc}
\hline\hline
ObsID  & $\Gamma$ & $f_{\rm sc}$ & $a_{*}$ & $\dot{M}$ & $f$ & $l$ & $\chi^{2} / \rm dof$ & Reduced-$\chi^{2}$ \\ 
& & & & ($10^{18}\,\rm {g/s}$) & & & & \\ \hline
P011466108801 & $1.97 _{-0.08} ^{+0.07}$ & $0.017  _{-0.005 } ^{+0.004 }$ & $0.08  _{-0.03 } ^{+0.04 }$ & $2.80 _{-0.08} ^{+0.07}$ &  1.65 & 0.137 & 296.4/433 & 0.68 \\
P011466108802 & $1.87 _{-0.08} ^{+0.08}$ & $0.018  _{-0.004 } ^{+0.005 }$ & $0.06  _{-0.04 } ^{+0.04 }$ & $2.84 _{-0.08} ^{+0.09}$ &  1.65 & 0.137 & 284.7/433 & 0.66 \\
P011466108901 & $1.77 _{-0.10} ^{+0.10}$ & $0.0059 _{-0.0017} ^{+0.003 }$ & $0.16  _{-0.019} ^{+0.014}$ & $2.52 _{-0.04} ^{+0.04}$ &  1.65 & 0.130 & 301.0/433 & 0.70 \\
P011466109001 & $1.80 _{-0.09} ^{+0.07}$ & $0.0072 _{-0.0019} ^{+0.003 }$ & $0.13  _{-0.03 } ^{+0.015}$ & $2.64 _{-0.04} ^{+0.06}$ &  1.65 & 0.133 & 287.0/433 & 0.66 \\
P011466109101 & $2.08 _{-0.08} ^{+0.08}$ & $0.021  _{-0.006 } ^{+0.006 }$ & $0.04  _{-0.04 } ^{+0.04 }$ & $2.83 _{-0.08} ^{+0.09}$ &  1.65 & 0.135 & 298.2/433 & 0.69 \\
P011466109102 & $1.93 _{-0.10} ^{+0.12}$ & $0.011  _{-0.004 } ^{+0.009 }$ & $0.12  _{-0.03 } ^{+0.03 }$ & $2.66 _{-0.06} ^{+0.15}$ &  1.65 & 0.133 & 298.5/433 & 0.69 \\
P011466109201 & $1.88 _{-0.08} ^{+0.08}$ & $0.010  _{-0.003 } ^{+0.005 }$ & $0.12  _{-0.03 } ^{+0.02 }$ & $2.61 _{-0.06} ^{+0.07}$ &  1.65 & 0.131 & 308.5/433 & 0.71 \\
P011466109202 & $1.89 _{-0.15} ^{+0.09}$ & $0.012  _{-0.004 } ^{+0.005 }$ & $0.11  _{-0.03 } ^{+0.03 }$ & $2.62 _{-0.07} ^{+0.07}$ &  1.65 & 0.131 & 317.7/433 & 0.73 \\
P011466109301 & $1.78 _{-0.10} ^{+0.10}$ & $0.0047 _{-0.0012} ^{+0.0016}$ & $0.134 _{-0.016} ^{+0.012}$ & $2.56 _{-0.03} ^{+0.04}$ &  1.65 & 0.130 & 319.4/433 & 0.74 \\
P011466109401 & $1.71 _{-0.11} ^{+0.11}$ & $0.0035 _{-0.0009} ^{+0.0014}$ & $0.141 _{-0.012} ^{+0.012}$ & $2.50 _{-0.03} ^{+0.03}$ &  1.65 & 0.127 & 310.9/433 & 0.72 \\
P011466109501 & $1.80 _{-0.12} ^{+0.11}$ & $0.005  _{-0.0015} ^{+0.002 }$ & $0.129 _{-0.018} ^{+0.012}$ & $2.52 _{-0.03} ^{+0.04}$ &  1.65 & 0.127 & 222.5/433 & 0.51 \\
P011466109502 & $1.92 _{-0.14} ^{+0.18}$ & $0.007  _{-0.003 } ^{+0.007 }$ & $0.125 _{-0.04 } ^{+0.017}$ & $2.55 _{-0.04} ^{+0.09}$ &  1.65 & 0.128 & 268.1/433 & 0.62 \\
P011466109503 & $1.99 _{-0.18} ^{+0.14}$ & $0.010  _{-0.005 } ^{+0.008 }$ & $0.10  _{-0.05 } ^{+0.03 }$ & $2.61 _{-0.07} ^{+0.11}$ &  1.65 & 0.129 & 351.8/433 & 0.81 \\
P011466109601 & $1.95 _{-0.06} ^{+0.06}$ & $0.015  _{-0.003 } ^{+0.004 }$ & $0.07  _{-0.03 } ^{+0.03 }$ & $2.65 _{-0.06} ^{+0.06}$ &  1.65 & 0.129 & 270.4/433 & 0.62 \\
P011466109602 & $1.97 _{-0.07} ^{+0.07}$ & $0.015  _{-0.004 } ^{+0.004 }$ & $0.07  _{-0.03 } ^{+0.03 }$ & $2.64 _{-0.06} ^{+0.07}$ &  1.65 & 0.128 & 305.1/433 & 0.70 \\
P011466109701 & $1.99 _{-0.06} ^{+0.06}$ & $0.016  _{-0.004 } ^{+0.004 }$ & $0.06  _{-0.03 } ^{+0.03 }$ & $2.61 _{-0.06} ^{+0.07}$ &  1.65 & 0.126 & 295.4/433 & 0.68 \\
P011466109702 & $1.91 _{-0.12} ^{+0.07}$ & $0.012  _{-0.006 } ^{+0.004 }$ & $0.08  _{-0.03 } ^{+0.04 }$ & $2.58 _{-0.09} ^{+0.07}$ &  1.65 & 0.126 & 345.0/433 & 0.80 \\
P011466109801 & $1.85 _{-0.15} ^{+0.14}$ & $0.005  _{-0.002 } ^{+0.004 }$ & $0.109 _{-0.03 } ^{+0.014}$ & $2.45 _{-0.03} ^{+0.06}$ &  1.65 & 0.122 & 248.6/433 & 0.57 \\
P011466109802 & $1.99 _{-0.08} ^{+0.10}$ & $0.011  _{-0.003 } ^{+0.004 }$ & $0.07  _{-0.03 } ^{+0.03 }$ & $2.52 _{-0.06} ^{+0.06}$ &  1.65 & 0.122 & 286.4/433 & 0.66 \\
P011466109803 & $1.94 _{-0.18} ^{+0.15}$ & $0.008  _{-0.005 } ^{+0.004 }$ & $0.09  _{-0.03 } ^{+0.04 }$ & $2.47 _{-0.07} ^{+0.06}$ &  1.65 & 0.122 & 261.1/433 & 0.60 \\
P011466109901 & $1.77 _{-0.14} ^{+0.15}$ & $0.0030 _{-0.0010} ^{+0.0019}$ & $0.107 _{-0.016} ^{+0.009}$ & $2.42 _{-0.02} ^{+0.04}$ &  1.65 & 0.120 & 264.7/433 & 0.61 \\
P011466110001 & $1.9  _{-0.3 } ^{+0.3 }$ & $0.0015 _{-0.0007} ^{+0.0015}$ & $0.138 _{-0.010} ^{+0.009}$ & $2.30 _{-0.02} ^{+0.02}$ &  1.63 & 0.116 & 241.2/433 & 0.56 \\
P011466110101 & $1.9  _{-0.3 } ^{+0.3 }$ & $0.0015 _{-0.0007} ^{+0.0016}$ & $0.152 _{-0.012} ^{+0.011}$ & $2.21 _{-0.03} ^{+0.03}$ &  1.63 & 0.113 & 249.1/433 & 0.58 \\
P011466110301 & $1.8  _{-0.4 } ^{+0.4 }$ & $0.0011 _{-0.0006} ^{+0.0018}$ & $0.149 _{-0.012} ^{+0.011}$ & $2.14 _{-0.02} ^{+0.03}$ &  1.63 & 0.109 & 294.7/433 & 0.68 \\
P011466110302 & $2.3  _{-0.4 } ^{+0.4 }$ & $0.0030 _{-0.0018} ^{+0.004 }$ & $0.129 _{-0.02 } ^{+0.012}$ & $2.18 _{-0.03} ^{+0.04}$ &  1.63 & 0.110 & 268.7/433 & 0.62 \\
P011466110401 & $2.5  _{-0.4 } ^{+0.4 }$ & $0.004  _{-0.002 } ^{+0.005 }$ & $0.175 _{-0.0016}^{+0.014}$ & $2.06 _{-0.04} ^{+0.04}$ &  1.64 & 0.107 & 274.9/433 & 0.63 \\
P011466110402 & $2.6  _{-0.6 } ^{+0.4 }$ & $0.005  _{-0.004 } ^{+0.008 }$ & $0.194 _{-0.03 } ^{+0.017}$ & $1.98 _{-0.03} ^{+0.05}$ &  1.63 & 0.104 & 268.4/433 & 0.62 \\
P011466110701 & $2.3  _{-0.4 } ^{+0.3 }$ & $0.0028 _{-0.0017} ^{+0.004 }$ & $0.161 _{-0.019} ^{+0.013}$ & $2.03 _{-0.03} ^{+0.04}$ &  1.63 & 0.105 & 228.7/433 & 0.53 \\
P011466110801 & $2.0  _{-0.5 } ^{+0.5 }$ & $0.0014 _{-0.0008} ^{+0.003 }$ & $0.151 _{-0.016} ^{+0.012}$ & $2.04 _{-0.03} ^{+0.03}$ &  1.63 & 0.104 & 300.3/433 & 0.69 \\
P011466110901 & $2.9  _{-0.4 } ^{+0.3 }$ & $0.008  _{-0.005 } ^{+0.007 }$ & $0.158 _{-0.02 } ^{+0.018}$ & $1.94 _{-0.03} ^{+0.04}$ &  1.63 & 0.099 & 239.4/433 & 0.55 \\
P011466110902 & $2.7  _{-0.7 } ^{+0.5 }$ & $0.006  _{-0.005 } ^{+0.011 }$ & $0.18  _{-0.04 } ^{+0.02 }$ & $1.89 _{-0.04} ^{+0.06}$ &  1.63 & 0.099 & 348.5/433 & 0.81 \\
P011466111001 & $2.6  _{-0.3 } ^{+0.3 }$ & $0.007  _{-0.003 } ^{+0.007 }$ & $0.18  _{-0.03 } ^{+0.02 }$ & $1.88 _{-0.03} ^{+0.05}$ &  1.63 & 0.098 & 231.6/433 & 0.53 \\
P011466111101 & $2.7  _{-0.4 } ^{+0.4 }$ & $0.005  _{-0.003 } ^{+0.007 }$ & $0.175 _{-0.02 } ^{+0.017}$ & $1.84 _{-0.03} ^{+0.04}$ &  1.63 & 0.095 & 268.3/433 & 0.62 \\
P011466111301 & $2.9  _{-0.6 } ^{+0.4 }$ & $0.008  _{-0.006 } ^{+0.006 }$ & $0.20  _{-0.03 } ^{+0.02 }$ & $1.75 _{-0.03} ^{+0.05}$ &  1.62 & 0.092 & 269.9/433 & 0.62 \\
P011466111401 & $3.3  _{-0.5 } ^{+0.8 }$ & $0.014  _{-0.009 } ^{+0.013 }$ & $0.16  _{-0.03 } ^{+0.03 }$ & $1.75 _{-0.04} ^{+0.05}$ &  1.62 & 0.090 & 275.8/433 & 0.64 \\
P011466111501 & $3.0  _{-0.4 } ^{+0.4 }$ & $0.012  _{-0.006 } ^{+0.008 }$ & $0.17  _{-0.03 } ^{+0.03 }$ & $1.69 _{-0.04} ^{+0.04}$ &  1.62 & 0.087 & 267.2/433 & 0.62 \\
P011466111601 & $2.8  _{-0.3 } ^{+0.3 }$ & $0.016  _{-0.006 } ^{+0.009 }$ & $0.148 _{-0.005} ^{+0.03 }$ & $1.67 _{-0.06} ^{+0.04}$ &  1.62 & 0.085 & 316.1/433 & 0.73 \\
P011466111701 & $2.3  _{-0.6 } ^{+0.5 }$ & $0.0015 _{-0.0011} ^{+0.003 }$ & $0.205 _{-0.017} ^{+0.010}$ & $1.522_{-0.019}^{+0.03}$ &  1.61 & 0.081 & 337.7/433 & 0.78 \\
P011466111801 & $2.5  _{-0.5 } ^{+0.5 }$ & $0.013  _{-0.007 } ^{+0.010 }$ & $0.16  _{-0.04 } ^{+0.04 }$ & $1.52 _{-0.06} ^{+0.06}$ &  1.61 & 0.078 & 387.5/433 & 0.90 \\
P011466111802 & $1.4  _{-0.4 } ^{+1.1 }$ & $0.0004 _{-0.0002} ^{+0.007 }$ & $0.23  _{-0.08 } ^{+0.04 }$ & $1.40 _{-0.06} ^{+0.11}$ &  1.61 & 0.076 & 312.3/365 & 0.86 \\
P011466111901 & $2.5  _{-0.6 } ^{+0.5 }$ & $0.003  _{-0.002 } ^{+0.005 }$ & $0.206 _{-0.03 } ^{+0.013}$ & $1.41 _{-0.02} ^{+0.05}$ &  1.60 & 0.075 & 288.7/433 & 0.67 \\
P011466112001 & $2.9  _{-0.5 } ^{+0.6 }$ & $0.010  _{-0.008 } ^{+0.010 }$ & $0.18  _{-0.03 } ^{+0.04 }$ & $1.40 _{-0.05} ^{+0.04}$ &  1.60 & 0.073 & 362.3/433 & 0.84 \\
P011466112101 & $3.2  _{-0.4 } ^{+0.4 }$ & $0.018  _{-0.008 } ^{+0.011 }$ & $0.14  _{-0.03 } ^{+0.03 }$ & $1.39 _{-0.04} ^{+0.04}$ &  1.60 & 0.071 & 319.4/433 & 0.74 \\
P011466112201 & $2.8  _{-0.4 } ^{+0.4 }$ & $0.017  _{-0.007 } ^{+0.009 }$ & $0.12  _{-0.04 } ^{+0.03 }$ & $1.38 _{-0.05} ^{+0.05}$ &  1.60 & 0.069 & 397.6/433 & 0.92 \\
P011466112301 & $2.6  _{-0.4 } ^{+0.4 }$ & $0.017  _{-0.007 } ^{+0.010 }$ & $0.10  _{-0.06 } ^{+0.06 }$ & $1.37 _{-0.07} ^{+0.08}$ &  1.60 & 0.068 & 349.8/398 & 0.88 \\
P011466112401 & $2.3  _{-0.5 } ^{+0.5 }$ & $0.013  _{-0.005 } ^{+0.008 }$ & $0.07  _{-0.03 } ^{+0.03 }$ & $1.21 _{-0.04} ^{+0.06}$ &  1.60 & 0.059 & 379.1/433 & 0.88 \\
P011466112402 & $2.0  _{-0.6 } ^{+0.6 }$ & $0.007  _{-0.003 } ^{+0.006 }$ & $0.13  _{-0.04 } ^{+0.04 }$ & $1.13 _{-0.05} ^{+0.05}$ &  1.57 & 0.057 & 326.7/391 & 0.84 \\
P011466112501 & $2.1  _{-0.4 } ^{+0.4 }$ & $0.008  _{-0.003 } ^{+0.004 }$ & $0.12  _{-0.02 } ^{+0.02 }$ & $1.12 _{-0.03} ^{+0.03}$ &  1.57 & 0.056 & 323.8/433 & 0.75 \\
P011466112601 & $2.2  _{-0.3 } ^{+0.3 }$ & $0.011  _{-0.003 } ^{+0.004 }$ & $0.10  _{-0.03 } ^{+0.02 }$ & $1.10 _{-0.03} ^{+0.03}$ &  1.57 & 0.054 & 355.0/433 & 0.82
     \\ \hline
    \end{tabular}%
    \end{center}
\end{table*}

\linespread{1.8}
\begin{table*}
    \caption{Fitting results for Model M4. $\Gamma$ is the photon index; $f_{\rm sc}$ is the scattering fraction; $a_*$ is the spin; $\dot{M}$ is the mass accretion rate in units of $10^{18}$~g/s; $R_{\rm ref}$ is the reflection fraction; $f$ is the spectral hardening factor; $l$ is the Eddington-scaled luminosity.}
    \label{tbl:M4}
    \begin{center}
    \begin{threeparttable}
    \begin{tabular}{cccccccccc}
    \hline\hline
    ObsID & $\Gamma$ & $f_{\rm sc}$ & $a_{*}$ & $\dot{M}$ & $R_{\rm ref}$ & $f$ & $l$ & $\chi^{2} / \rm dof$ & Reduced-$\chi^{2}$ \\ 
    & & & & ($10^{18}\, \rm {g/s}$) & & & & & \\ \hline 
 P011466108801 & $2.13 _{-0.04} ^{+0.05}$ & $0.0280 _{-0.0005} ^{+0.0014}$ & $0.064 _{-0.013} ^{+0.008}$ & $2.84 _{-0.03}  ^{+0.04}$ & $1.04 _{-0.10} ^{+0.11}$ & 1.65 & 0.137 & 762.5/867 & 0.87 \\
 P011466108802 & $2.06 _{-0.05} ^{+0.05}$ & $0.0259 _{-0.0018} ^{+0.0009}$ & $0.070 _{-0.007} ^{+0.016}$ & $2.82 _{-0.04}  ^{+0.03}$ & $0.99 _{-0.09} ^{+0.13}$ & 1.65 & 0.137 & 687.0/783 & 0.87 \\
 P011466108901 & $1.97 _{-0.05} ^{+0.05}$ & $0.0159 _{-0.0009} ^{+0.0010}$ & $0.138 _{-0.011} ^{+0.011}$ & $2.57 _{-0.03}  ^{+0.03}$ & $0.85 _{-0.13} ^{+0.13}$ & 1.64 & 0.130 & 635.7/762 & 0.83 \\
 P011466109001 & $2.04 _{-0.06} ^{+0.03}$ & $0.0194 _{-0.0010} ^{+0.0006}$ & $0.086 _{-0.010} ^{+0.014}$ & $2.73 _{-0.02}  ^{+0.03}$ & $0.97 _{-0.09} ^{+0.09}$ & 1.65 & 0.134 & 779.3/877 & 0.88 \\
 P011466109101 & $2.16 _{-0.05} ^{+0.05}$ & $0.0254 _{-0.0013} ^{+0.0016}$ & $0.058 _{-0.015} ^{+0.013}$ & $2.80 _{-0.05}  ^{+0.04}$ & $1.23 _{-0.3 } ^{+0.05}$ & 1.65 & 0.135 & 623.1/787 & 0.79 \\
 P011466109102 & $2.07 _{-0.04} ^{+0.03}$ & $0.0221 _{-0.0010} ^{+0.0009}$ & $0.070 _{-0.012} ^{+0.010}$ & $2.76 _{-0.04}  ^{+0.03}$ & $1.00 _{-0.11} ^{+0.10}$ & 1.65 & 0.134 & 640.5/807 & 0.79 \\
 P011466109201 & $2.02 _{-0.05} ^{+0.02}$ & $0.0207 _{-0.0012} ^{+0.0009}$ & $0.096 _{-0.014} ^{+0.014}$ & $2.66 _{-0.04}  ^{+0.04}$ & $0.90 _{-0.12} ^{+0.12}$ & 1.64 & 0.131 & 706.7/797 & 0.88 \\
 P011466109202 & $2.04 _{-0.03} ^{+0.06}$ & $0.0179 _{-0.0010} ^{+0.0012}$ & $0.114 _{-0.011} ^{+0.008}$ & $2.60 _{-0.03}  ^{+0.03}$ & $1.03 _{-0.13} ^{+0.15}$ & 1.64 & 0.130 & 726.1/805 & 0.90 \\
 P011466109301 & $2.05 _{-0.05} ^{+0.04}$ & $0.0165 _{-0.0012} ^{+0.0008}$ & $0.103 _{-0.016} ^{+0.010}$ & $2.63 _{-0.03}  ^{+0.02}$ & $0.94 _{-0.11} ^{+0.10}$ & 1.64 & 0.130 & 640.4/772 & 0.82 \\
 P011466109401 & $2.01 _{-0.05} ^{+0.05}$ & $0.0131 _{-0.0008} ^{+0.0008}$ & $0.110 _{-0.009} ^{+0.010}$ & $2.56 _{-0.03}  ^{+0.03}$ & $0.99 _{-0.13} ^{+0.13}$ & 1.64 & 0.127 & 641.3/780 & 0.82 \\
 P011466109501 & $2.04 _{-0.04} ^{+0.04}$ & $0.0133 _{-0.0006} ^{+0.0007}$ & $0.112 _{-0.004} ^{+0.003}$ & $2.55 _{-0.01}  ^{+0.02}$ & $1.02 _{-0.11} ^{+0.12}$ & 1.64 & 0.127 & 744.7/895 & 0.83 \\
 P011466109502 & $2.12 _{-0.06} ^{+0.06}$ & $0.0173 _{-0.0013} ^{+0.0013}$ & $0.091 _{-0.014} ^{+0.014}$ & $2.62 _{-0.04}  ^{+0.03}$ & $1.08 _{-0.17} ^{+0.17}$ & 1.64 & 0.129 & 605.7/776 & 0.78 \\
 P011466109503 & $2.13 _{-0.07} ^{+0.07}$ & $0.0208 _{-0.0016} ^{+0.0019}$ & $0.075 _{-0.017} ^{+0.018}$ & $2.66 _{-0.02}  ^{+0.05}$ & $0.88 _{-0.2 } ^{+0.19}$ & 1.64 & 0.130 & 562.9/674 & 0.83 \\
 P011466109601 & $2.07 _{-0.04} ^{+0.04}$ & $0.0204 _{-0.0008} ^{+0.0009}$ & $0.079 _{-0.010} ^{+0.007}$ & $2.62 _{-0.02}  ^{+0.02}$ & $0.94 _{-0.10} ^{+0.10}$ & 1.64 & 0.128 & 709.1/877 & 0.80 \\
 P011466109602 & $2.04 _{-0.02} ^{+0.05}$ & $0.0207 _{-0.0007} ^{+0.0010}$ & $0.075 _{-0.012} ^{+0.007}$ & $2.62 _{-0.02}  ^{+0.03}$ & $0.99 _{-0.10} ^{+0.11}$ & 1.64 & 0.128 & 719.0/814 & 0.88 \\
 P011466109701 & $2.08 _{-0.04} ^{+0.04}$ & $0.0239 _{-0.0010} ^{+0.0011}$ & $0.051 _{-0.014} ^{+0.010}$ & $2.63 _{-0.02}  ^{+0.04}$ & $0.94 _{-0.10} ^{+0.11}$ & 1.64 & 0.126 & 669.2/841 & 0.79 \\
 P011466109702 & $2.06 _{-0.06} ^{+0.04}$ & $0.0214 _{-0.0011} ^{+0.0009}$ & $0.053 _{-0.009} ^{+0.015}$ & $2.62 _{-0.04}  ^{+0.02}$ & $1.01 _{-0.13} ^{+0.12}$ & 1.64 & 0.126 & 740.5/834 & 0.88 \\
 P011466109801 & $2.07 _{-0.05} ^{+0.05}$ & $0.0133 _{-0.0007} ^{+0.0008}$ & $0.086 _{-0.012} ^{+0.010}$ & $2.49 _{-0.03}  ^{+0.03}$ & $1.07 _{-0.14} ^{+0.08}$ & 1.64 & 0.122 & 707.2/869 & 0.81 \\
 P011466109802 & $2.08 _{-0.03} ^{+0.08}$ & $0.0147 _{-0.0008} ^{+0.0009}$ & $0.078 _{-0.007} ^{+0.013}$ & $2.50 _{-0.03}  ^{+0.02}$ & $1.01 _{-0.15} ^{+0.14}$ & 1.64 & 0.122 & 719.6/847 & 0.84 \\
 P011466109803 & $2.10 _{-0.06} ^{+0.06}$ & $0.0128 _{-0.0009} ^{+0.0010}$ & $0.091 _{-0.012} ^{+0.011}$ & $2.47 _{-0.01}  ^{+0.03}$ & $1.17 _{-0.16} ^{+0.17}$ & 1.64 & 0.121 & 729.9/844 & 0.86 \\
 P011466109901 & $2.09 _{-0.05} ^{+0.04}$ & $0.0110 _{-0.0003} ^{+0.0007}$ & $0.075 _{-0.008} ^{+0.010}$ & $2.48 _{-0.02}  ^{+0.02}$ & $0.96 _{-0.14} ^{+0.14}$ & 1.64 & 0.121 & 677.8/855 & 0.79 \\
 P011466110001 & $2.33 _{-0.10} ^{+0.09}$ & $0.0073 _{-0.0011} ^{+0.0012}$ & $0.126 _{-0.010} ^{+0.009}$ & $2.31 _{-0.02}  ^{+0.02}$ & $1.3  _{-0.3 } ^{+0.3 }$ & 1.63 & 0.116 & 673.6/837 & 0.80 \\
 P011466110101 & $2.09 _{-0.10} ^{+0.10}$ & $0.0042 _{-0.0007} ^{+0.0008}$ & $0.151 _{-0.009} ^{+0.009}$ & $2.20 _{-0.02}  ^{+0.02}$ & $1.3  _{-0.3 } ^{+0.3 }$ & 1.63 & 0.112 & 657.1/808 & 0.81 \\
 P011466110301 & $2.25 _{-0.12} ^{+0.12}$ & $0.0054 _{-0.0012} ^{+0.0014}$ & $0.134 _{-0.011} ^{+0.009}$ & $2.16 _{-0.02}  ^{+0.02}$ & $1.5  _{-0.4 } ^{+0.5 }$ & 1.63 & 0.109 & 618.4/795 & 0.77 \\
 P011466110302 & $2.46 _{-0.14} ^{+0.16}$ & $0.0071 _{-0.0016} ^{+0.002 }$ & $0.129 _{-0.011} ^{+0.009}$ & $2.17 _{-0.02}  ^{+0.02}$ & $1.3  _{-0.4 } ^{+0.4 }$ & 1.63 & 0.109 & 698.9/817 & 0.85 \\
 P011466110401 & $2.72 _{-0.13} ^{+0.14}$ & $0.013  _{-0.003 } ^{+0.003 }$ & $0.142 _{-0.010} ^{+0.011}$ & $2.09 _{-0.02}  ^{+0.02}$ & $1.2  _{-0.3 } ^{+0.3 }$ & 1.63 & 0.106 & 679.1/830 & 0.81 \\
 P011466110402 & $2.59 _{-0.14} ^{+0.14}$ & $0.012  _{-0.003 } ^{+0.004 }$ & $0.166 _{-0.012} ^{+0.011}$ & $2.02 _{-0.03}  ^{+0.03}$ & $1.8  _{-0.5 } ^{+0.7 }$ & 1.62 & 0.104 & 632.5/764 & 0.82 \\
 P011466110701 & $2.46 _{-0.13} ^{+0.14}$ & $0.0088 _{-0.0017} ^{+0.002 }$ & $0.145 _{-0.008} ^{+0.009}$ & $2.06 _{-0.02}  ^{+0.02}$ & $1.2  _{-0.3 } ^{+0.4 }$ & 1.62 & 0.104 & 685.3/878 & 0.78 \\
 P011466110801 & $2.37 _{-0.14} ^{+0.15}$ & $0.0061 _{-0.0017} ^{+0.002 }$ & $0.138 _{-0.010} ^{+0.010}$ & $2.05 _{-0.02}  ^{+0.02}$ & $1.7  _{-0.5 } ^{+0.7 }$ & 1.62 & 0.104 & 693.9/789 & 0.87 \\
 P011466110901\tnote{a} & $3.1  _{-0.2 } ^{+0.2 }$ & $0.0164 _{-0.005}  ^{+0.005 }$ & $0.152 _{-0.010} ^{+0.018}$ & $1.95 _{-0.03}  ^{+0.02}$ & $1.3  _{-0.3 } ^{+0.3 }$ & 1.63 & 0.100 & 714.0/828 & 0.86 \\
 P011466110902 & $2.69 _{-0.2 } ^{+0.3 }$ & $0.010  _{-0.003 } ^{+0.005 }$ & $0.174 _{-0.015} ^{+0.015}$ & $1.90 _{-0.03}  ^{+0.03}$ & $1.4  _{-0.6 } ^{+0.7 }$ & 1.62 & 0.098 & 574.1/675 & 0.85 \\
 P011466111001 & $2.53 _{-0.12} ^{+0.13}$ & $0.013  _{-0.002 } ^{+0.003 }$ & $0.158 _{-0.010} ^{+0.011}$ & $1.91 _{-0.02}  ^{+0.02}$ & $1.3  _{-0.3 } ^{+0.4 }$ & 1.62 & 0.098 & 634.6/828 & 0.76 \\
 P011466111101\tnote{a} & $2.96 _{-0.14} ^{+0.16}$ & $0.0112 _{-0.004}  ^{+0.005 }$ & $0.136 _{-0.015} ^{+0.011}$ & $1.90 _{-0.02}  ^{+0.03}$ & $2.7  _{-0.8 } ^{+0.9 }$ & 1.63 & 0.096 & 654.0/787 & 0.83 \\
 P011466111301 & $2.90 _{-0.20} ^{+0.2 }$ & $0.014  _{-0.004 } ^{+0.004 }$ & $0.181 _{-0.010} ^{+0.015}$ & $1.77 _{-0.03}  ^{+0.02}$ & $1.5  _{-0.5 } ^{+0.6 }$ & 1.61 & 0.092 & 606.2/763 & 0.79 \\
 P011466111401\tnote{a} & $3.2  _{-0.3 } ^{+0.4 }$ & $0.0122 _{-0.006}  ^{+0.009 }$ & $0.161 _{-0.014} ^{+0.014}$ & $1.74 _{-0.03}  ^{+0.03}$ & $2.1  _{-0.7 } ^{+1.2 }$ & 1.62 & 0.090 & 635.9/783 & 0.81 \\
 P011466111501\tnote{a} & $3.2  _{-0.3 } ^{+0.3 }$ & $0.0141 _{-0.007}  ^{+0.007 }$ & $0.166 _{-0.011} ^{+0.014}$ & $1.69 _{-0.03}  ^{+0.02}$ & $2.2  _{-0.5 } ^{+1.0 }$ & 1.62 & 0.087 & 677.0/778 & 0.87 \\
 P011466111601 & $2.55 _{-0.16} ^{+0.18}$ & $0.008  _{-0.003 } ^{+0.004 }$ & $0.187 _{-0.013} ^{+0.014}$ & $1.59 _{-0.03}  ^{+0.02}$ & $2.1  _{-0.6 } ^{+1.2 }$ & 1.61 & 0.083 & 611.4/721 & 0.84 \\
 P011466111701 & $2.74 _{-0.19} ^{+0.2 }$ & $0.010  _{-0.003 } ^{+0.005 }$ & $0.175 _{-0.014} ^{+0.013}$ & $1.56 _{-0.02}  ^{+0.03}$ & $1.7  _{-0.6 } ^{+0.8 }$ & 1.61 & 0.081 & 658.2/710 & 0.92 \\
 P011466111801 & $2.33 _{-0.2 } ^{+0.2 }$ & $0.011  _{-0.004 } ^{+0.006 }$ & $0.177 _{-0.018} ^{+0.023}$ & $1.49 _{-0.04}  ^{+0.03}$ & $1.7  _{-0.8 } ^{+1.3 }$ & 1.61 & 0.077 & 551.7/632 & 0.87 \\
 P011466111802\tnote{a} & $2.54 _{-0.15} ^{+0.08}$ & $0.0030 _{-0.002 } ^{+0.008 }$ & $0.14  _{-0.03 } ^{+0.03 }$ & $1.50 _{-0.05}  ^{+0.05}$ & $13   _{-10  } ^{+30  }$\tnote{b} & 1.61 & 0.076 & 435.3/463 & 0.94 \\
 P011466111901 & $2.46 _{-0.16} ^{+0.19}$ & $0.010  _{-0.003 } ^{+0.004 }$ & $0.136 _{-0.015} ^{+0.012}$ & $1.49 _{-0.03}  ^{+0.03}$ & $2.0  _{-0.7 } ^{+1.0 }$ & 1.60 & 0.075 & 571.4/694 & 0.82 \\
 P011466112001 & $2.59 _{-0.17} ^{+0.2 }$ & $0.007  _{-0.005 } ^{+0.006 }$ & $0.167 _{-0.018} ^{+0.010}$ & $1.41 _{-0.02}  ^{+0.03}$ & $3    _{-5   } ^{+2   }$ & 1.60 & 0.073 & 624.5/665 & 0.94 \\
 P011466112101\tnote{a} & $3.1  _{-0.2 } ^{+0.3 }$ & $0.0219 _{-0.007}  ^{+0.012 }$ & $0.136 _{-0.018} ^{+0.017}$ & $1.41 _{-0.03}  ^{+0.03}$ & $1.4  _{-0.4 } ^{+0.5 }$ & 1.60 & 0.071 & 560.9/684 & 0.82 \\
 P011466112201\tnote{a} & $2.63 _{-0.12} ^{+0.07}$ & $0.0116 _{-0.005}  ^{+0.005 }$ & $0.102 _{-0.010} ^{+0.014}$ & $1.40 _{-0.02}  ^{+0.03}$ & $3.1  _{-0.7 } ^{+1.5 }$ & 1.60 & 0.070 & 612.9/617 & 0.99 \\
 P011466112301\tnote{a} & $2.97 _{-0.18} ^{+0.15}$ & $0.0209 _{-0.007}  ^{+0.008 }$ & $0.07  _{-0.03 } ^{+0.03 }$ & $1.40 _{-0.05}  ^{+0.04}$ & $2.4  _{-0.5 } ^{+0.5 }$ & 1.59 & 0.068 & 542.6/515 & 1.05 \\
 P011466112401 & $2.54 _{-0.2 } ^{+0.3 }$ & $0.014  _{-0.006 } ^{+0.010 }$ & $0.106 _{-0.03 } ^{+0.016}$ & $1.17 _{-0.02}  ^{+0.04}$ & $1.7  _{-0.8 } ^{+1.4 }$ & 1.57 & 0.058 & 601.1/655 & 0.91 \\
 P011466112402 & $2.03 _{-0.08} ^{+0.15}$ & $0.004  _{-0.004 } ^{+0.003 }$ & $0.136 _{-0.020} ^{+0.024}$ & $1.12 _{-0.03}  ^{+0.03}$ & $3.3  _{-0.9 } ^{+4.0 }$ & 1.57 & 0.057 & 409.9/500 & 0.81 \\
 P011466112501 & $1.93 _{-0.17} ^{+0.17}$ & $0.0092 _{-0.0019} ^{+0.002 }$ & $0.119 _{-0.013} ^{+0.011}$ & $1.12 _{-0.02}  ^{+0.02}$ & $1.2  _{-0.5 } ^{+0.3 }$ & 1.57 & 0.056 & 611.4/721 & 0.84 \\
 P011466112601 & $2.36 _{-0.14} ^{+0.16}$ & $0.011  _{-0.003 } ^{+0.004 }$ & $0.113 _{-0.015} ^{+0.011}$ & $1.08 _{-0.01}  ^{+0.02}$ & $1.9  _{-0.6 } ^{+0.8 }$ & 1.57 & 0.054 & 618.2/694 & 0.89

\\ \hline

\end{tabular}
\begin{tablenotes}
    \item[\textbf{Notes.}]
    \item[a] The CONSTANT of ME is allowed to vary from 0.9 to 1.1 due to the poor constraint caused by the statistical fluctuation in data.
    \item[b] poor constraint due to short exposure of LE.
    \end{tablenotes}
    \end{threeparttable}
    
\end{center}
\par
\end{table*}

\bsp
\label{lastpage}

\end{document}